\documentclass{article}

\usepackage[english]{babel}

\usepackage[letterpaper,top=2cm,bottom=2cm,left=3cm,right=3cm,marginparwidth=1.75cm]{geometry}
\usepackage[version=3]{mhchem} 
\usepackage{float}
\usepackage{xcolor}
\usepackage{bm}
\usepackage{amssymb}
\usepackage{amsmath}
\usepackage{subfig}
\usepackage{graphicx}
\usepackage{comment}
\usepackage{physics}
\usepackage{algorithm}
\usepackage{algorithmic}
\usepackage{cleveref}
\usepackage{float} 
\usepackage{subcaption}
\usepackage{authblk}
\usepackage{svg}
\usepackage{array}
\usepackage{tikz}
\usepackage[doi=false,isbn=false,url=false,eprint=false]{biblatex}

\bibliography{ref.bib}

\title{Learning Density Functionals to Bridge Particle and Continuum Scales}
\author[1]{Edoardo Monti}
\author[2]{Peter Yatsyshin} 
\author[3]{Konstantinos Gkagkas} 
\author[4]{Andrew B. Duncan}

\affil[1]{Department of Mathematics, Imperial College London, London SW7 2AZ, United Kingdom}
\affil[2]{The Alan Turing Institute, London NW1 2DB, United Kingdom}
\affil[3]{Advanced Technology Division, Toyota Motor Europe NV/SA, Technical Center, Hoge Wei 33B, 1930 Zaventem, Belgium}
\affil[4]{Department of Mathematics, Imperial College London, London SW7 2AZ, United Kingdom}

\DeclareMathOperator*{\argmin}{argmin}   

\begin{document}
\maketitle

\begin{abstract}
    Predicting interfacial thermodynamics across molecular and continuum scales remains a central challenge in computational science. Classical density functional theory (cDFT) provides a first-principles route to connect microscopic interactions with macroscopic observables, but its predictive accuracy depends on approximate free-energy functionals that are difficult to generalize. Here we introduce a physics-informed learning framework that augments cDFT with neural corrections trained directly against molecular-dynamics data through adjoint optimization. Rather than replacing the theory with a black-box surrogate, we embed compact neural networks within the Helmholtz free-energy functional, learning local and nonlocal corrections that preserve thermodynamic consistency while capturing missing correlations. Applied to Lennard-Jones fluids, the resulting augmented excess free-energy functional quantitatively reproduces equilibrium density profiles, coexistence curves, and surface tensions across a broad temperature range, and accurately predicts contact angles and droplet shapes far beyond the training regime. This approach combines the interpretability of statistical mechanics with the adaptability of modern machine learning, establishing a general route to learned thermodynamic functionals that bridge molecular simulations and continuum-scale models.
\end{abstract}
\section{Introduction}\label{sec:intro}
Predicting macroscopic interfacial behavior from microscopic interactions remains a central challenge in computational physics and materials science.  Phenomena, such as wetting, capillary transport and phase coexistence depend sensitively on molecular correlations at nanometer scales.  On the other hand, technological applications, including coatings and porous transport \cite{coating, wang2023_porous_wetting}, nanofluidic devices ~\cite{zhang_nanofluidic} and water management in fuel cells~\cite{YANG20241284, pefc}, require continuum models capable of describing the effective behavior over microns or millimeters.  
\\\\
Bridging these scales requires characterizations of matter which are simultaneously thermodynamically consistent, computationally tractable, and grounded in molecular physics.  Classical density functional theory (cDFT)~\cite{evans1992fundamentals, cdftreview, 11_ddft} offers a first-principles route to building this connection. Rooted in statistical mechanics, cDFT expresses the equilibrium structure of an inhomogeneous fluid in terms of a Helmholtz Free Energy functional (HFE)~\cite{evans1992fundamentals} of the one-body density $\rho(\bm{r})$. It captures molecular layering and interfacial thermodynamics at modest cost, avoiding the statistical noise and size limitations of molecular dynamics (MD).
However, the exact functional form of the excess Helmholtz Free Energy $F_{ex}[\rho]$ is unknown and must be approximated~\cite{roth2010_fmt,tarazona2008_fmt_review}.  Traditional perturbative or semi-empirical closures, such as Barker–Henderson~\cite{BH1} or weighted-density formulations~\cite{wda, mwda}, are capable of reproducing simple systems, such as Lennard-Jones (LJ) fluids, but lose quantitative accuracy when transferred to more complex fluids. Optimizing complex multiphysics devices requires macroscopic models that faithfully represent the underlying physical processes. In practice, this means constitutive relations must be improved beyond empirical corrections, which often lack predictive power.
\\\\
Addressing this inverse problem has been a central theme in modern statistical-mechanical literature~\cite{roth2010_fmt,tarazona2008_fmt_review}. Semi-Empirical HFE models exist, but they remain confined to simple fluids (hard-spheres, LJ). Recent work has shown that machine learning can be systematically integrated into liquid-state theory and classical density functional theory by learning effective free-energy or correlation functionals while preserving the underlying variational structure~\cite{12_ml_review_cdft,3_neur_cdft_schimdt_bulk,2_neur_cdft_evans_ml_bulk,7_neur_cdft_from_inter_potential,8_neur_cdft_ddft,14_neur_cdft_computing_mu_from_sim_data, 9_neur_cdft_matching_c2}. These approaches have demonstrated that ML-augmented functionals can reproduce exact results in analytically tractable settings, such as one-dimensional hard-rod systems~\cite{percus_hr,5_neur_cdft_schimdt_percus}, as well as more complex fluids, including polar~\cite{6_neur_cdft_appl_co2}, ionic~\cite{16_neur_cdft_ionic_fluids}, and multicomponent mixtures~\cite{13_neur_cdft_bulk_binary_mixtures}.
However, purely data-driven approaches such as Machine Learning (ML) surrogates face intrinsic challenges. Neural operators and, in general, regression deep learning models are usually limited to interpolation and lack physical consistency and extrapolative reliability, severely limiting their ability to generalize into unseen temperature and material regimes~\cite{extrapolation_issues,Zhu_2023}.   In this context, direct surrogate modeling of the HFE by means of a neural operator may be impractical, because it acts as a functional over an infinite-dimensional domain of spatial density functions. Reliable extrapolation would therefore require extensive sampling across this vast domain, which would be again computationally prohibitive. 
Here we propose a physics-informed ML framework that unites the rigor of statistical mechanics with the adaptability of modern machine learning.
The philosophy behind this approach is based on the premise that physical laws and constraints are inherently embedded in the observable data and can be exploited to guide the development of more robust and accurate models. 
Rather than replacing $F_\text{ex}$ with an unconstrained neural model, we retain its established physical decomposition into short-range repulsion, long-range attraction, and nonlocal packing contributions, and introduce learned corrections that act as low-dimensional, thermodynamically coherent functions of the local state variables $(\rho, T)$.
Each correction is represented by a small-footprint neural network trained via adjoint optimization, which enforces the Euler–Lagrange equilibrium conditions of cDFT at every iteration. 
This strong physics-based inductive bias improves training efficiency and enables robust learning under realistic conditions of data scarcity.
This yields an augmented $F_\text{ex}$ that preserves physical interpretability while adapting quantitatively to molecular-scale data.
\\\\
Applied to the test case of the well-known LJ fluid, the learned functional reproduces MD-derived density profiles, phase coexistence curves, and liquid–gas surface tensions with quantitative fidelity across temperatures and system size, while maintaining thermodynamic interpretability and transferable behavior under new geometries and external potentials.
As the corrections depend only on local thermodynamic variables, the trained model generalizes naturally to higher-dimensional geometries, such as droplets and patterned surfaces, providing a scalable bridge from atomistic simulations to continuum-scale predictions.
\\\\
Among the macroscopic observables that encode interfacial thermodynamics, the contact angle plays a particularly critical role~\cite{Yuan2013}.
It quantifies the balance of surface tensions at the three-phase contact line and thus governs wetting, adhesion, and transport across a vast range of systems: controlling condensation and frosting on engineered surfaces, liquid infiltration in porous electrodes and membranes, lubrication and coating performance, and even capillary-driven assembly in nanofabrication.
Because the contact angle depends on the local thermodynamic state, surface chemistry, and molecular-scale structure, it cannot be treated as a fixed material constant ~\cite{ASKARIPOUR2019118705, wettability_tmd_dependence, DUCHEMIN2021215}.
Traditional continuum models rely on empirical correlations or single-value measurements that fail under changing temperature, composition, or confinement.
Atomistic simulations via Molecular Dynamics (MD) (e.g. via LAMMPS \cite{LAMMPS}) or Monte Carlo (MC) can provide accurate data for systematic mapping of these dependencies,  but remain computationally prohibitive.   Moreover, these methods are bounded to nano-scale systems, which doesn't immediately allow for direct extrapolation and application of the results to macroscopic models.   By learning the underlying free-energy functional itself, our approach enables quantitative, transferable prediction of equilibrium contact angles across geometries and operating conditions—achieving, within a single variational framework, the fidelity of molecular simulation and the efficiency of continuum modelling.
\\\\
Our approach establishes a general paradigm for learning thermodynamic functionals: embedding differentiable neural components within variational physical theories to combine interpretability, data efficiency, and predictive reach within a hybrid physics-ML framework.    A key feature of this calibration is its
theoretical independence from the external potential: the HFE is an intrinsic property of the fluid rather than of the confining geometry. Consequently, the same functional can be efficiently applied to different physical setups, such as varying wall materials or geometries,  without the need for retraining. Beyond wetting, the same framework can inform next-generation closures for complex mixtures, electrochemical interfaces, and non-equilibrium multiphase systems, offering a new route to seamless multiscale modeling in computational science.
\\\\
Our work makes four main contributions.
\begin{enumerate}
    \item We introduce a hybrid statistical–machine-learning framework that augments classical density functional theory with data-driven, thermodynamically biased neural corrections, enabling accurate free-energy modeling without sacrificing interpretability.
\item We formulate an adjoint-based optimization procedure that enforces the Euler–Lagrange equilibrium conditions of cDFT during training, providing an efficient route to differentiable calibration from molecular data.
\item We demonstrate that the resulting augmented excess Helmholtz free energy (eHFE) reproduces molecular dynamics reference results for the well-known Lennard–Jones fluid across wide thermodynamic ranges, yielding accurate density profiles, phase coexistence curves, surface tensions, and thermodynamic-dependent contact angles.
\item We show that the learned functional maintains transferability and extrapolative power, generalizing from one-dimensional calibration data to higher-dimensional geometries and unseen conditions.
\end{enumerate}

The remainder of the paper is organized as follows.
Section 2 introduces the theoretical background of classical density functional theory in both canonical and grand-canonical formulations, together with standard approximations for the excess free energy $F_\text{ex}$.
Section 3 presents the inverse-design formulation and the construction of the physics-informed neural corrections to $F_\text{ex}$ within the cDFT framework.
Section 4 details the adjoint-based training methodology and reports calibration results for planar wetting against molecular dynamics data, including quantitative validation of phase behavior and surface tension.
Section 5 extends the approach to three-dimensional droplet geometries to predict contact angles and assess transferability.
Section 6 analyzes the calibrated machine-learning corrections and provides insight into the unexpectedly strong extrapolation capabilities of the augmented model.
Finally, Section 7 summarizes the key findings and outlines future directions toward multicomponent fluids, non-equilibrium systems, and integration with continuum-scale solvers.
\\\\

\section{Theoretical Background on Classical Density Functional Theory}
\label{sec:Theory}

In this section, the theory underlying this work and the terminology used are briefly reviewed, with emphasis on its  grand canonical and canonical formulation, for consistent comparison with MD data, and the commonly used approximations for the HFE.

\paragraph{Variational Formulation in Grand Canonical Ensemble}
In classical DFT, we describe an inhomogeneous fluid by its one-body number density function $\rho(\bm r)$, which gives the average local density at a point $\bm r \in D \subset \mathbb{R}^3$.  Let $\Omega$ be the grand potential, for a spatially inhomogeneous single-component system, described by $\rho(\bm r)$, we have
\vspace{0.5em}
\begin{equation}
    \label{eq:OmegaFunctional}
    \begin{aligned}
    \Omega([\rho], \mu, T) \;=\; F([\rho], T) \;+\; \int_D d\bm r \,\rho(\bm r)\!\left(V_{\text{ext}}(\bm r)-\mu\right),
    \end{aligned}
\end{equation}
where $\mu$ and $T$ are the chemical potential and the temperature respectively, $V_{\text{ext}}$ is the external potential acting on the system, and $F$ is the HFE. The functional $F$ contains the intrinsic thermodynamics of the fluid, which is independent of the external field $V_{\text{ext}}$. 
$F$ can be decomposed into ideal and excess contributions~\cite{evans1992fundamentals}:
\vspace{0.5em}
\begin{equation}
    \label{eq:HFEdecomposition}
    \begin{aligned}
        F([\rho], T) \;=\; F_{\text{id}}([\rho], T) \;+\; F_{\text{ex}}([\rho], T).
    \end{aligned}
\end{equation}
From now on, for the sake of lighter notation, the explicit thermodynamic potentials' dependence on $T$ and $\mu$ is omitted if not necessary.
The excess term $F_{\text{ex}}$ is not known exactly, and the ideal contribution has an analytical form:
\vspace{0.5em}
\begin{equation}
    \begin{aligned}
    \label{eq:Fid}
        F_{\text{id}}[\rho] \;=\; \cfrac{1}{\beta}\int_D \;\rho(\bm r)\!\left[\ln\!\big(\Lambda^3 \rho(\bm r)\big)-1\right] \; d\bm r ,
    \end{aligned}
\end{equation}
where $\beta = 1/(k_B T)$ is the reciprocal temperature, $k_B$ is the Boltzmann constant and $\Lambda$ is the thermal De Broglie wavelength~\cite{mcquarrie2000statistical}. In practice, $\Lambda$ may be subsumed into $\mu$.
In the grand canonical ensemble, at given $\beta$ and $\mu$, the equilibrium density $\rho_\text{eq}(\bm r)$ may be obtained by minimizing $\Omega[\rho]$~\cite{evans1992fundamentals}: 
\vspace{0.5em}
\begin{equation}
    \label{eq:gc_minimization}
    \begin{aligned}
        \rho_\text{eq}(\bm r; \beta,\mu) = \argmin_{\rho(\bm r)} \Omega([\rho(\bm r)], \beta,\mu ),
    \end{aligned}
\end{equation}
leading to the Euler–Lagrange (EL) equation:

\vspace{0.5em}
\begin{equation}
    \label{eq:EL-muVT}
    \begin{aligned}
        \rho_\text{eq} : h_\text{gc}([\rho_\text{eq}], \beta,\mu) \;=\;0,
    \end{aligned}
\end{equation}
with:
\vspace{0.5em}
\begin{equation}
    \label{eq:h-muVT}
    h_\text{gc}([\rho], \beta,\mu) \;=\; \rho(\bm r) - \Lambda^{-3} 
    \exp \!\left[ \beta \!\left( \mu - \frac{\delta F_{\text{ex}}[\rho]}{\delta \rho(\bm r)} - V_{\text{ext}}(\bm r) \right) \right] .
\end{equation}
%
\paragraph{Canonical Ensemble}
To facilitate direct comparison with MD data, we constrain the number of particles using a Lagrange multiplier. This effectively corresponds to canonical ensemble in our case, because average $N$ is large and its fluctuations are negligible~\cite{White2000}:
\vspace{0.5em}
\begin{equation}
    \label{eq:c_minimization}
    \begin{aligned}
        \rho_\text{eq}(\bm r; \beta, N) &\;:\; h_\text{gc}([\rho_\text{eq}(\bm r)]; \beta,\mu )] \;=\; 0, \\
        \textrm{subject to}\;\; & g([\rho_\text{eq}(\bm r)], N) \;=\; 0,
    \end{aligned}
\end{equation}
where $g=0$ is an integral constraint ensuring that the total number of particles equals $N$:
\vspace{0.5em}
\begin{equation}
    \label{eq:c_constraint}
    g([\rho],N) \;=\; \int_D  \rho(\bm r)\, d\bm r - N \;=\; 0.   
\end{equation}
This constraint can be analytically resolved and a corresponding canonical ensemble counterpart of Eq.~\eqref{eq:EL-muVT} can be obtained:
\vspace{0.5em}
\begin{equation}
    \label{eq:EL-NVT}
    \begin{aligned}
        \rho_\text{eq} : h_\text{c}([\rho_\text{eq}], \beta,N) \;=\;0,
    \end{aligned}
\end{equation}
with:
\vspace{0.5em}
\begin{equation}
    \label{eq:h-NVT}
    h_\text{c}([\rho], \beta,N) \;=\; \rho(\bm r) \;-\; \frac{N \,\exp \left[ -\beta\!\left(\,\tfrac{\delta F_{\text{ex}}[\rho]}{\delta \rho( \bm r)} \;+\; V_{\text{ext}}(\bm r)\,\right)  \right]  }
       {\displaystyle \int_D
        \exp \left[ -\beta\!\left(\,\tfrac{\delta F_{\text{ex}}[\rho]}{\delta \rho( \bm r')} \;+\; V_{\text{ext}}(\bm r')\,\right)  \right] d\bm r'} .
\end{equation}
This canonical cDFT formulation allows for direct comparison with data obtained from $NVT$ molecular dynamics simulations, ensuring a one-to-one correspondence between the two frameworks.
\subsection{Excess Helmholtz Free Energy}
\label{sec:excessHFE_standard}

The predictive capability of classical DFT critically relies on the accuracy of the excess-over-ideal Helmholtz free energy functional, $F_\text{ex}[\rho]$, which accounts for contributions arising from interparticle interactions. $F_\text{ex}$ is generally unknown and must be modeled.

\paragraph{The Barker--Henderson Perturbation Theory.} 
The Barker-Henderson (BH) theory~\cite{BH1} expands the fluid's free energy by treating strong repulsive forces as the reference system and weak attractive forces as small perturbations. The pair potential $u(r)$ typically can be decomposed as follows:
\begin{equation}
    \label{eq:h-NVT}
    u(r) = u^\text{rep}(r) + u^\text{att}(r),
\end{equation}
where $u_\text{rep}(r)$ represents a strongly repulsive interaction and $u_\text{att}(r)$ the weaker attractive tail. This allows us to approximate $F_\text{ex}$ using perturbation theory, modeling the dominant short-range packing effects as a reference hard-sphere fluid, while incorporating the long-range cohesive forces through a perturbative expansion.
For LJ fluids, the pair potential is given by
\begin{equation}
    u_{\text{LJ}}(r;\; \bm p) = 4\varepsilon\left[\left(\frac{\sigma}{r}\right)^{12}
    - \left(\frac{\sigma}{r}\right)^{6}\right],
    \label{eq:lj126}
\end{equation}
where the LJ parameters $\bm p=[\varepsilon,\,\sigma]$ respectively control attractive well depth and the effective particle diameter.
In this work, the 
interaction is truncated and shifted at a cutoff radius $r_c = 2.5\sigma$, i.e. the pair potential is shifted so that\
$u_{\text{LJ}}(r)=0$ for $r\ge r_c$. This ensures continuity of the potential at the cutoff.
According to the BH prescription, the LJ repulsive part is given by
\begin{equation}
    u^{\text{rep}}_\text{LJ}(r) =
    \begin{cases}
        u_{\text{LJ}}(r) - u_{\text{LJ}}(r_m), & r < r_m, \\[4pt]
        0, & r \ge r_m, 
    \end{cases}
\end{equation}
where $r_m = 2^{1/6}\sigma$ is the location of the $u_{\text{LJ}}(r)$ minimum. The corresponding repulsive contribution to the free energy is then approximated by that of a hard-sphere fluid with temperature-dependent diameter $\sigma_{\text{bh}}(T)$:
\begin{equation}
\sigma_\text{bh} = \int_0^{+\infty}(1-e^{-\beta u_\text{rep}(r)}) dr.
\end{equation}
On the other hand, the attractive tail 
\begin{equation}
    u^{\text{att}}_\text{LJ}(r; \; \bm p) =
    \begin{cases}
        u_{\text{LJ}}(r_m; \; \bm p), & r < r_m, \\[4pt]
        u_{\text{LJ}}(r; \; \bm p), & r \ge r_m,
    \end{cases}
\end{equation}
is treated perturbatively. This separation allows to split the contributions to $F_\text{ex}$ in two distinct terms, $F_\text{rep}$ describing short-range packing effects and $F_\text{att}$ -- the long-range attractive interactions: 
\vspace{0.5em}
\begin{equation}
\label{eq:bh_rep_att}
    F_\text{ex}[\rho] = F_{\text{rep}}[\rho] + F_{\text{att}}[\rho].
\end{equation}
A popular choice for the repulsive contribution $F_{\text{rep}}$ is the phenomenological Carnahan–Starling (CS) free energy, which corresponds to the equation of state (EOS) for a hard-sphere fluid~\cite{cs}:

\begin{equation}
\label{eq:F_rep_lda}
F_{\text{rep}}[\rho] \;=\; \int_D  f_{\text{rep}}\big(\rho(\bm r)\big)\;d\bm r ,
\end{equation}
with $f_{\text{rep}}$ being the excess free energy per unit volume and given by~\cite{ cs_free_energy}:
\begin{equation}
\label{fcs}
   f_{\text{rep}}(\rho) \;=\; \cfrac{\rho}{\beta} \, \frac{4\eta(\rho) - 3\eta(\rho)^{2}}{\big(1-\eta(\rho)\big)^{2}},
\end{equation}
where $\eta \in [0, 1]$ is the packing fraction:
\begin{equation}
    \eta(\bm x) \;=\; \cfrac{\pi}{6} \rho(\bm x) \sigma_\text{bh}^3,
\end{equation}
representing the fraction of total volume occupied by the fluid.
\paragraph{Weighted Density Approximation.}
\label{subsubsec:wda}
Introducing spatial dependence of density into Eq.~\eqref{fcs} is non-trivial. The simple approach of replacing $\rho$ with $\rho({\bm r})$, known as Local Density Approximation (LDA)~\cite{lda_error}, is unable to capture inter-particle correlations responsible for near-wall layering effects. The simplest way to account for correlations is by introducing a weighted density $\bar{\rho}(\bm{r})$ into $f_{\text{rep}}$:
\begin{equation}
\label{eq:weighted_density}
    \bar{\rho}(\bm{r}) = \int_D \rho(\bm{r}')\, w(|\bm{r}-\bm{r}'|; \bar{\rho})\,d\bm{r}',
\end{equation}
where $w$ is a short-ranged kernel reflecting the spatial extent of repulsive correlations.
This is known as the Weighted Density Approximation (WDA)~\cite{wda, mwda}. The  repulsive part of the free-energy functional is then expressed as
\begin{equation}
    F_\text{rep}[\rho] = \int_D  f_\text{rep}(\bar{\rho}(\bm r))\,d\bm r.
\end{equation}
To recover the bulk limit, the weighting function must satisfy the normalization condition~\cite{mwda}:
\begin{equation}
    \int_0^{+\infty} 4\pi r^2  w(r; \bar{\rho})\, dr= 1.
\end{equation}
The non-locality of WDA allows $\rho({\bf r})$ of inhomogeneous fluid to exhibit the proper oscillatory structure near interfaces, while preserving bulk thermodynamic properties. 
More advanced formulations, such as the Modified WDA (MWDA)~\cite{mwda} and the Fundamental Measure Theory (FMT)~\cite{roth2010_fmt}, build upon the same principle using more rigorous geometrical weighting schemes.

\paragraph{$F_\text{att}$ for LJ Fluids.}
\label{}
The BH expansion beyond the 0-th order repulsive term, accounts for the long-range attractive part of the intermolecular interactions. 
The perturbation series for the attractive free energy is formally written as:
\begin{equation}
    F_\text{att} =  F_\text{att}^{(1)} + F_\text{att}^{(2)} + ...,
\end{equation}
and is typically truncated after the first term. The resulting first-order contribution is given by the following expression~\cite{theory_of_inhomogenous_fluids}:
\begin{equation}
\label{eq:1st_ord_BH}
    F_\text{att}[\rho] = \frac{1}{2} \iint_{D \times D}  \rho(\bm r)\rho(\bm r') \; g_\text{hs}(\bm r, \bm r';[\rho]) \, u_\text{LJ}^\text{att}(|\bm r-\bm r'|; \; \bm p) \, d\bm r d\bm r',
\end{equation}
where $g_\text{hs}$ is the pair-correlation functional of the reference HS fluid (the 0th-order system), which depends non-locally on the density field. Although the temperature does not appear explicitly in the above expression, it enters implicitly through $g_\text{hs}$.
Evaluating Eq.~\eqref{eq:1st_ord_BH} requires knowing $g_\text{hs}$. In principle, $g_\text{hs}$ can be obtained from the Ornstein–Zernike equation for the reference HS fluid~\cite{theory_of_inhomogenous_fluids}. In practice, the Mean-Field (MF) approximation is widely used, where $g_\text{hs}$ is effectively set to unity.

\section{Learning the Helmholtz Free Energy Functional}
\label{sec:InverseProblem}

While the mean-field framework discussed above correctly captures the essential physics of inhomogeneous fluids, quantitative agreement with experiments and even MD simulations remains a challenge. Much effort is presently directed towards improving theoretical prescriptions for $F_{\mathrm{ex}}$~\cite{tarazona2008_fmt_review,wda,mwda,roth2010_fmt}. We are offering a hybrid machine-learning physics-constrained framework which uses fluid-state theory as a necessary inductive bias for ML. More precisely, we are advocating against a standard brute-force approach which attempts to directly learn $F_\text{ex}$ as a data-driven surrogate $F_{\mathrm{ex},\theta}$. This would require learning the following map, which is defined over an infinite-dimensional functional domain:
\begin{equation}
    \label{fullfunc}
     \underbrace{F_{\mathrm{ex},\theta}([\rho], T):
    (\mathcal{X}\times\mathbb{R}) \to \mathbb{R}}_{\text{infinite-dimensional, nonlocal}}
\qquad 
\mathcal{X} = \{\rho(\bm r):D\to\mathbb{R}_+\}.
\end{equation}
On the surface, Neural Operators (NOs) seem to offer an ML model capable of learning the functional map in Eq.~\eqref{fullfunc}~\cite{neural_operator}, albeit being computationally expensive to train and requiring a consistent amount of training data. However, the main problem with NOs is that they are only applicable in regression tasks, where extrapolation capability outside of the training manifold is not expected~\cite{extrapolation_issues,Zhu_2023}. As a result, such models typically suffer from poor generalization when applied outside the training domain.
%
%
We note that established statistical-mechanical theories may provide us with appropriate inductive bias, reducing the challenging learning task in Eq.~\eqref{fullfunc} to a much simpler task of learning several corrections ${\phi_\theta}^{(k)}$ to the approximate mean-field cDFT described in the previous section. The learning task \eqref{fullfunc} is then reduced from a regression over infinite-dimensional set \(\mathcal{X}\) to regression over two-dimensional set $(\rho, T)\in \mathcal A\subset\mathbb{R}^2$:
\begin{equation}
    \underbrace{{\phi_\theta}^{(k)}(\rho, T):
    \mathcal A \to \mathbb{R}}_{\text{two-dimensional, local}}.
\end{equation}
At a given temperature, each \({\phi_\theta}^{(k)}\) only depends on the local value of \(\rho \in \mathbb{R}^+\). Therefore, \({\phi_\theta}^{(k)}\)'s input space can be efficiently sampled from a limited set of higher-dimensional data points \(\rho(\bm {r}) \in \mathcal{X}\), which span the desired density range. We demonstrate that the dimensionality reduction offered by the properly chosen inductive bias dramatically reduces sampling complexity and enhances generalization, while  maintaining interpretability and the physical consistency. 
%
%
Compared to pure surrogate modeling, our physics-biased approach contains fewer trainable parameters and achieves greater efficiency, enabling robust learning under realistic conditions of data scarcity. Furthermore, the model does not depend on the spatial domain of the underlying physical system, allowing generalization to unseen geometries. In what follows, we describe our models for ${\phi_\theta}^{(k)}$.

\subsection{Augmented Helmholtz Free Energy}
\label{sec:DataDrivenHFE}

We follow the BH decomposition in \eqref{eq:bh_rep_att}, but introduce parameterized corrections for the repulsive and attractive terms. In this way we preserve the physical separation between short-range repulsion, long-range attraction, and correlation-driven phenomena, such as particle layering. 


\subsubsection{Correction to Bulk Thermodynamics.}\label{subsec:corr_1_rep}

To correctly capture the bulk of the LJ fluid, we augment the CS repulsive free-energy density with the following data-driven scaler: 
\begin{equation}
    \bar f_{\text{rep}}(\rho) \;=\; f_\text{cs}(\rho)
    \big( 1 + {\phi_{\theta}^{(1)}}(\rho, \beta ;\; \bm \theta_\text{cs}) \big),
\end{equation}
where ${\phi_{\theta}^{(1)}}:\mathcal A \to \mathbb R$ is a function of the local thermodynamic variables $\rho$ and $\beta$, represented as a Neural Network (NN) with trainable parameters $\bm \theta_\text{cs}$. In the limit $\rho\to0$, we should recover ideal fluid. This is enforced by the following condition:
\begin{equation}
    \frac{d}{d\rho} \left[{\bar f}_\text{rep} (\rho)\right]_{\rho \to 0} = 0. 
\end{equation}
In practice, the above condition states that $\phi_{\theta}^{(1)}$ and its first derivative must remain finite, since the Carnahan--Starling term decays quadratically as $\rho \!\to\! 0$.

\subsubsection{Nonlocal Correction to Capture the Fluid Correlation Structure.}\label{subsec:corr_2_wda}

By making the WDA kernel in Eq.~\eqref{eq:weighted_density} learnable, we aim to improve the accuracy of capturing the short-range repulsive correlations. For simplicity, we do not consider density dependence of the WDA kernel. This follows established approaches \cite{wda, roth2010_fmt}.
Since NN architectures tend to smear out high-frequency oscillations~\cite{nn_freq_bias}, we parametrize $w(r)$ in Fourier space. Therefore, the convolution in Eq.~\eqref{eq:weighted_density} reduces to a product:
\begin{equation}
    \hat{\bar\rho}(\bm k)
    \;=\; \hat\rho(\bm k)\, \hat w(\bm k),
\end{equation}
where a hat denotes the Fourier transform operator $\mathcal{F}$. 
The Fourier modes ${\hat \phi_{\theta}^{(2)}}: \mathbb{R}^+ \to \mathbb{R}$, parameterized by $\bm\theta_w$, represent the kernel in Fourier space, having $\hat {\bar w} = \hat \phi_{\theta}^{(2)}$ . The weighted density is now computed by the following expression:
\begin{equation}
    \bar\rho(\bm r;\;\bm\theta_w)
    \;=\;
    \mathcal{F}^{-1}\!\left[\hat\rho\, {\hat\phi_{\theta}^{(2)}}\right].
\end{equation}
To make our model less sensitive to noise and to unphysical long-ranged fluid-fluid interaction, we limit the extent of repulsive correlations by constraining $\phi_{\theta}^{(2)}$.
Concretely, we start from the Fourier-space kernel $\hat\phi_{\theta}^{(2)}(k)$, transform it to real space, and apply a Gaussian damping factor $\exp\!\left[-\frac{r^{2}}{2\sigma_c^{2}}\right]$, with $\sigma_c= 3$ a phenomenologically chosen damping  scale. This smoothly suppresses correlations for $r\gg \sigma_c$ while preserving the regularity of the kernel. The filtered kernel is then transformed back to Fourier space.
Additionally, to maintain the correct bulk limit, we further constrain $\phi_{\theta}^{(2)}$ to have a unitary norm in the real domain:
\begin{equation}
    \int_0^{+\infty} 4\pi r^2 \phi_{\theta}^{(2)}(r;\; \bm \theta_w)\,d r = 1.
\end{equation}
which is equivalent to constraining the kernel’s zero-frequency mode to unity in the Fourier domain, $\hat\phi_{\theta}^{(2)}(0;\;\bm\theta_w)=1$. 

The resulting $\bar F_\text{rep}$ is given by the following expression:
\begin{equation}
    \bar F_\text{rep}\left([\rho] ;\; \bm\theta_\text{cs}, \bm\theta_w \right)
    \;=\;
    \int_D 
    \bar f_\text{rep}\!\big(\bar\rho(\bm r)\big) \, d\bm r.
\end{equation}

\subsubsection{Corrections to the Attractive Functional} \label{subsec:corr_3_att}
Each LJ parameter $p_i \in \bm p$ is augmented with temperature-dependent NN corrections ${\phi_{\theta,i}^{(3)}}:\mathbb R^+ \to \mathbb R$:
\begin{equation}
    {p_\theta}_i (\beta;\bm{{\theta}_\text{att}}) 
    \;=\; p_i
    \big( 1 + {\phi_{\theta,i}^{(3)}}(\beta ;\; \bm \theta_\text{att}) \big).
\label{eq:phi3}
\end{equation}
We further introduce a learnable structure function $\phi_{\theta}^{(4)}:\mathbb{R}^+ \to \mathbb{R}$ to go beyond the mean-field approximation in Eq.~\eqref{eq:1st_ord_BH}:
\begin{equation}
    {\bar g} (r;\;\bm{{\theta}}_g) 
    \;=\; 1 + \phi^{(4)}_\theta(r;\;\bm{{\theta}}_g)
\end{equation}
Similarly to the learnable WDA kernel, we do not consider density-dependence of ${\bar g}$. With these corrections,  Eq.~\eqref{eq:1st_ord_BH} takes the following form:
\begin{equation}
    \bar F_\text{att}\left([\rho];\; \bm \theta_\text{att} \right) 
    \;=\;
    \frac{1}{2}
    \iint_{D\times D} 
    \rho(\bm r)\rho(\bm r')\, \bar g \left(|\bm r-\bm r'|;\;\bm{{\theta}}_g \right)
    u_\text{att}\!\left(|\bm r-\bm r'|;\; \bm p_\theta(\beta ;\; \bm \theta_\text{att})\right) \, d\bm r d\bm r'.
\end{equation}

In this respect, by introducing temperature dependence into the augmented LJ potential, we are allowing our ML model the flexibility to capture the appropriate temperature-driven trends, without the need to approximate solution to the Ornstein-Zernicke equation.

\begin{figure}[H]
    \centering
    \includegraphics[scale=0.25
    ]{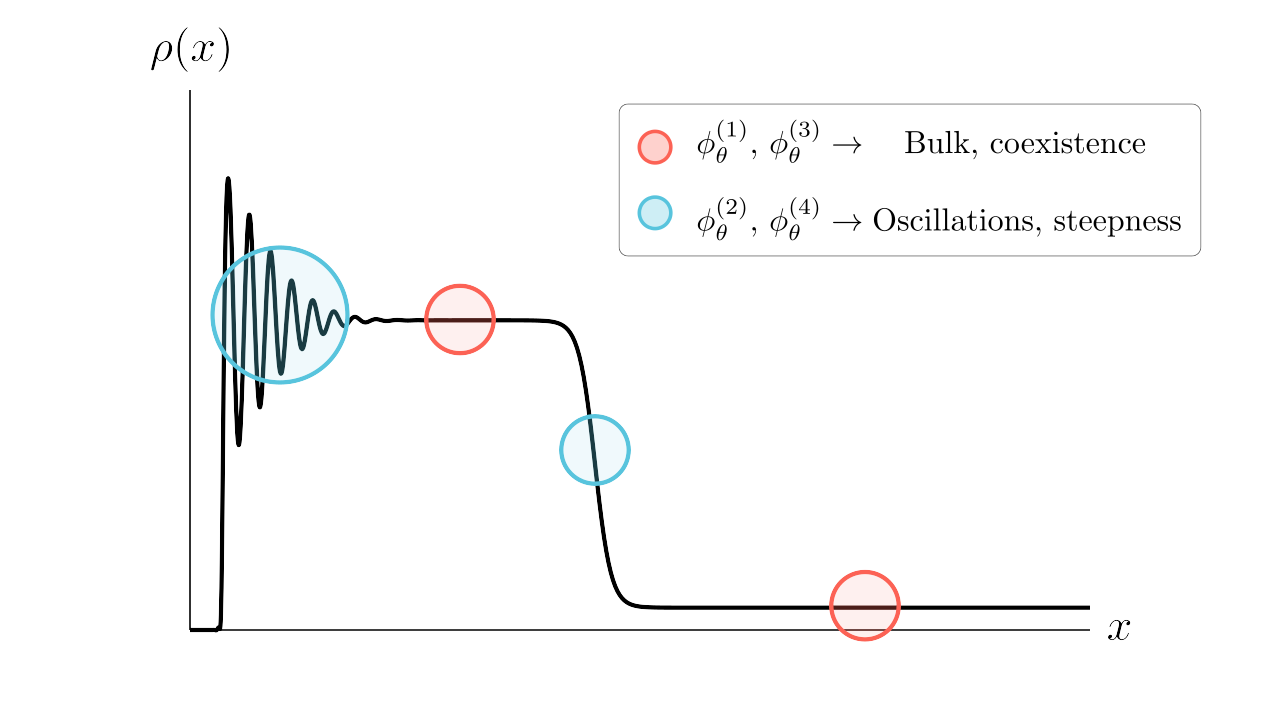}
    \caption{Relationship between the ML corrections to the physics-based DFT approximation and the features of the density profile of the adsorbed fluid. The correction terms $\phi_\theta^{(1, 3)}$ allows us to capture the bulk binodal, resulting in the correct liquid and gas plateaus of the density profile. 
    The corrections $\phi_\theta^{(2, 4)}$ allow us to capture the correct wall-liquid layering and the slope of the wall-vapor interface.
    }
    \label{fig:corr_contribution}
\end{figure}

In problems of adsorption, the corrections ${\phi_\theta}^{(k)}$ to the augmented functional $\bar F_\text{ex}= \bar F_\text{rep}+\bar F_\text{att}$ that we introduced can be traced back to the well-known features of the fluid density profile: near-wall layering, liquid-gas interface and near-constant plateaus at values of bulk coexisting densities. Figure~\ref{fig:corr_contribution} shows a typical density profile of an inhomogeneous fluid in contact with an attractive wall, identifying the correction terms we introduced with the profile features.

\section{Training and Validation on Planar Wetting}\label{sec:results
}
\label{sec:results}
We train the physics-constrained ML DFT functional using MD data for fluid adsorption on a planar wall. The resulting model functional not only correctly captures adsorption across a range of temperatures, but also recovers the bulk binodal, including the critical temperature of the LJ fluid. There is significant computational benefit to this protocol, because it only requires us to solve the DFT problem in one spatial dimension. Furthermore, we later demonstrate that our trained model generalizes very well to adsorption in three spatial dimensions. 

\subsection{Molecular Dynamics Reference Data and Baseline Accuracy}
\label{subsec:MDplanar}

\begin{figure}[hbt!]
    \centering
    \subfloat[]{\includegraphics[width=0.25\textwidth, clip, trim={0.0cm 0.0cm 1.0cm 0.5cm}]{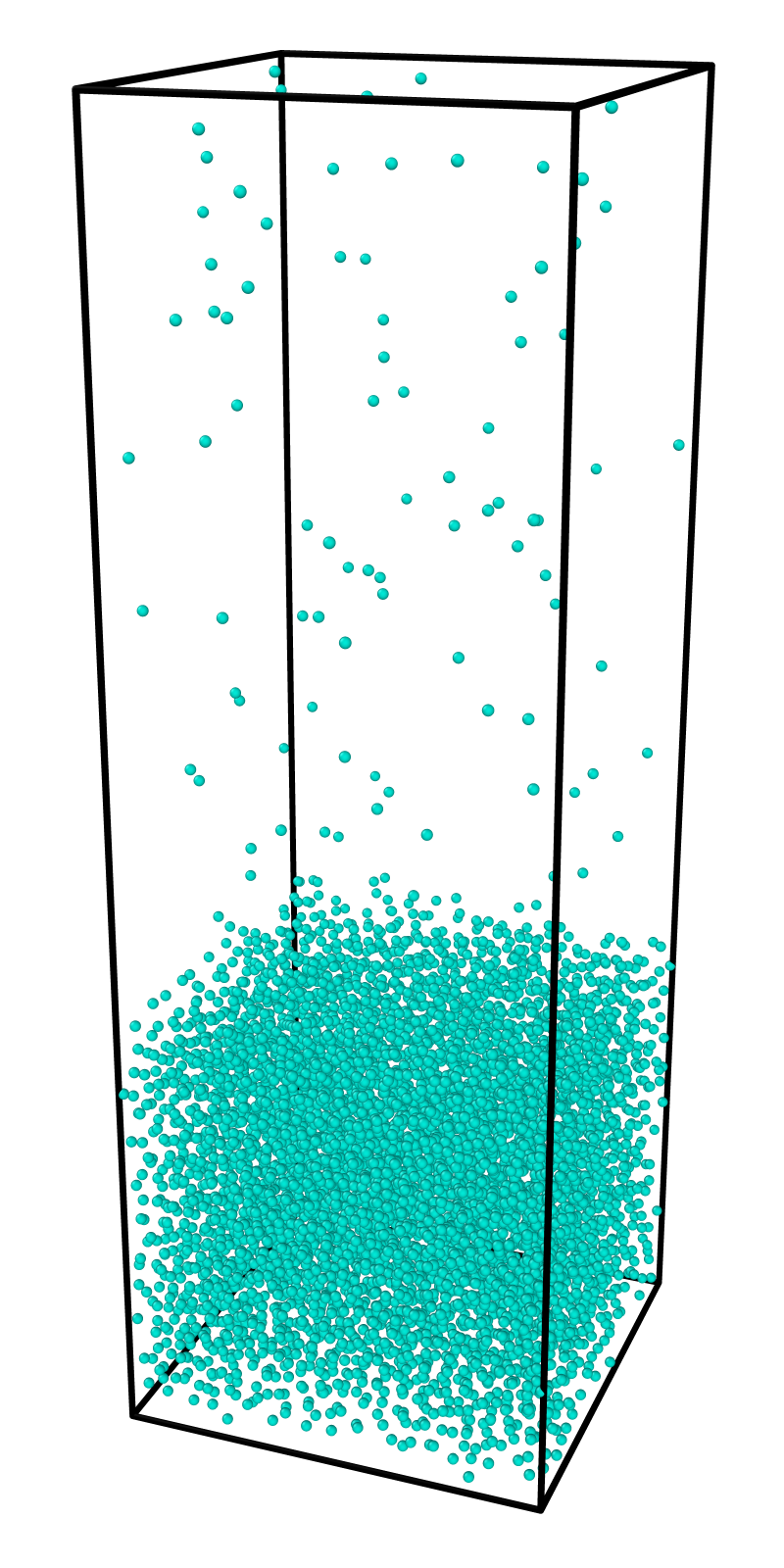}}
    \hspace{0cm}
    \subfloat[]{\includegraphics[width=0.25\textwidth, clip, trim={0.0cm 0.0cm 1.0cm 0.5cm}]{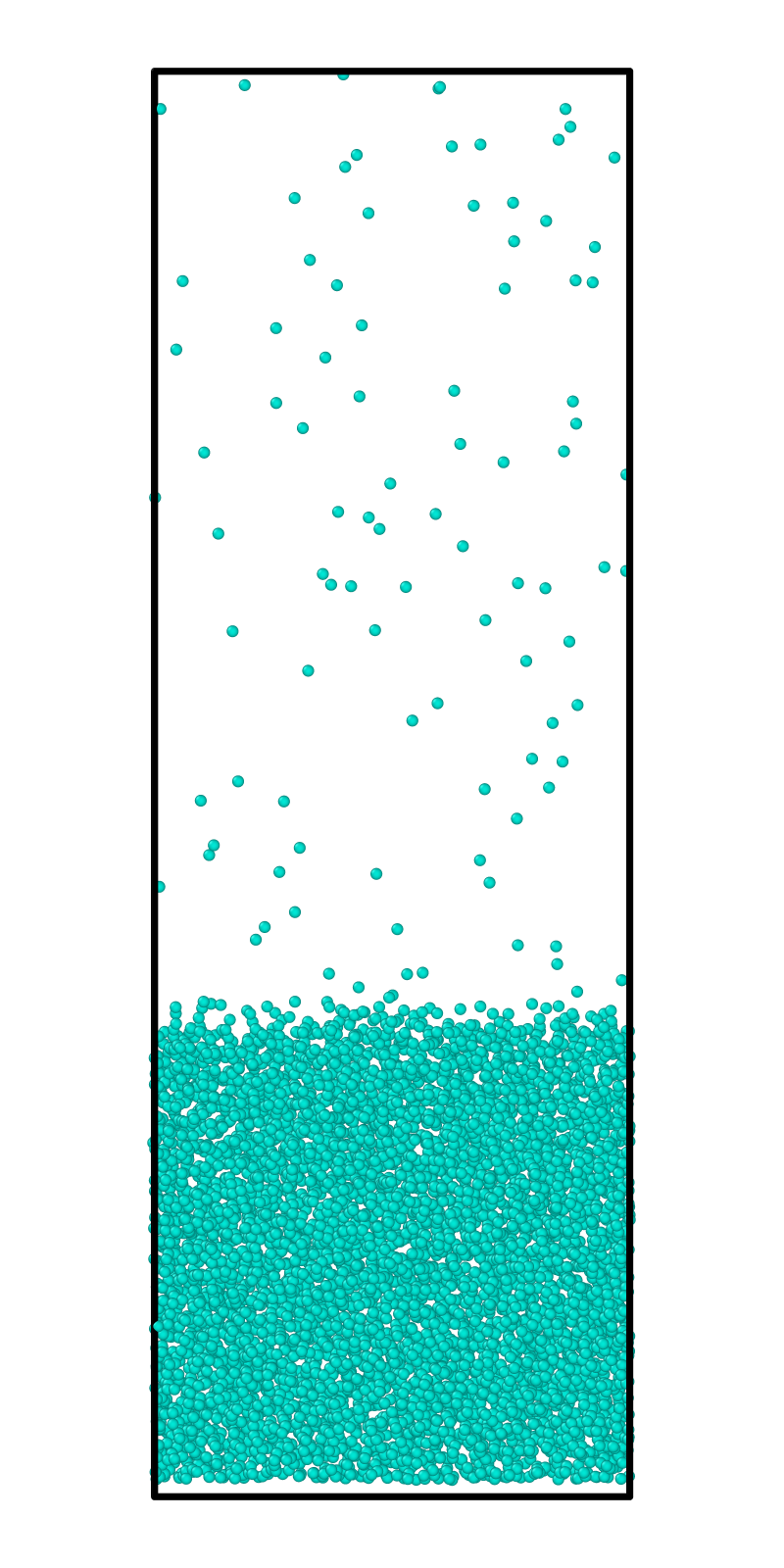}}
\caption{Illustrative example of the MD data showing liquid, adsorbed on the planar wall, in contact with its saturated vapor. Limited number of such simulations at different temperatures were used to train the augmented DFT: (a) perspective view and (b) frontal view. An equilibration run is performed to relax the system to its equilibrium state. After equilibrium is reached, a production run is carried out to collect the statistical data for analysis}
\label{fig:md_sim_box}
\end{figure}

We generated a dataset for the training and validation of $\bar F_\text{ex}$,, respectively  $\mathcal D_t$ and  $\mathcal D_\text{ref}$, by performing MD simulations of planar wetting in the $NVT$ ensemble with a Langevin thermostat~\cite{langevin_thermo}. All simulations were carried out using the LAMMPS package~\cite{LAMMPS}. The simulation domain is a three-dimensional box of size $L_x=60\sigma$ and $L_y=L_z=20\sigma$, as illustrated in Figure~\ref{fig:md_sim_box}.
The fluid–fluid interactions were given by the truncated and shifted LJ12--6 potential in Eq.~\eqref{eq:lj126} with $r_c=2.5$. The MD fluid model must match the cDFT one. 
We imposed an integrated LJ potential $V_{\text{ext}}$ at the bottom boundary ($x=x_0$) to mimic a semi-infinite wall located at $x<x_0$, interacting with the fluid via a fluid-wall LJ potential:
\begin{equation}
    V_{\text{ext}}(x;\;\bm p_w) 
    = \varepsilon_{w}\left[
    \frac{2}{15}\!\left(\frac{\sigma_{w}}{x}\right)^{9}
    - \left(\frac{\sigma_{w}}{x}\right)^{3}
    \right],
    \label{eq:lj93}
\end{equation}
where $\bm p_w=[\varepsilon_w, \sigma_w]$ are the wall--fluid LJ parameters.
At the top of the box, a reflective boundary condition was applied, while periodicity in the ${y}$- and ${z}$-directions was used to enforce planar symmetry.
For a given domain $D$ of volume $V$, the thermodynamic state of the fluid is specified by the particle number $N$ and the temperature $T$. Additionally, we need to provide the parameters $\bm p_w$ of the external potential. Thus, one data point refers to the adsorption density profile $\rho^{(i,j,k)}(x)$ specified by the triplet $(N^{(i)}_d,\,T^{(j)}_d,\,\bm p^{(k)}_w)$. The training dataset and testing dataset are constructed by using two different values of the parameters $\bm p_w$, allowing us to assess predictive performance on previously unseen wall configurations. The detailed dataset structure is explained in the Appendix A. $\bar F_\text{ex}$ was trained on five data points and validated on fifty data points, showing remarkable generalization capabilities. 
%
%

\subsection{Adjoint Optimization for Physics-Constrained Learning}
\label{subsec:TrainingOptimization}
Model parameters $\bm \theta = [\bm\theta_{cs}, \bm\theta_w, \bm\theta_{att}, \bm\theta_g]$ are calibrated by minimizing an objective functional $J$. The regularized $L_2$ norm of the relative error between the density field $\rho(x;\bm\theta)$, obtained from the augmented cDFT, and the training data $\rho_t(x)\in \mathcal D_t$, obtained from MD, is used to quantify the model's discrepancy from the target data:
\begin{equation}
    \label{eq:J}
    J([\rho(\bm{\theta})], \bm{\theta}) 
    = \frac{1}{2} \int_D  
    \left( \rho(\bm{\theta}) - \rho_t\right)^{\!2}\,  d x
    \,\Big/\int_D  
    \rho_t^{2} \, d x
    + \alpha_\theta \|\bm{\theta}\|^2,
\end{equation}
where we omit the spatial dependence to lighten the notation. The damping parameter  $\alpha_\theta=10^{-6}$ was empirically tuned to stabilize training and prevent overfitting~\cite{bengio12}. Scaling by $\int_D\rho_t^2\,dx$ is necessary when dealing with different temperatures, to account for the differences in density magnitude across the temperature range.

An alternative approach to $\bar F_\text{ex}$ could attempt to calibrate $\bm\theta$ by matching the higher-order structural information, such as $c^{(2)} = -\beta {\delta^2 \bar F_{\mathrm{ex}}([\rho];\bm\theta)}/ {\left(\delta \rho(\bm r_1)\, \delta \rho(\bm r_2)\right)}$, with its MD counterpart. This approach would enable direct gradient computation through auto-differentiation, but is only practical for bulk systems and very sensitive to noise in the data. Since we are interested in adsorption, the form of Eq.~\eqref{eq:J} is preferable as it directly deals with the inhomogeneous system. 
We train $\bar F_\text{ex}$ by solving the following constrained optimization problem:
\begin{equation}
\label{eq:Optimization}
\begin{aligned}
    &\bm{\theta}^\star 
    =  \argmin_{\bm{\theta}}
    \sum_{j} J\!\left(\left[\rho(\bm{\theta})\right], \bm{\theta}\,\big|\, \mathcal D_t^{(j)} \right), \\[4pt]
    &\text{subject to} \quad h_\text{c}\!\left([\rho],\bm{\theta}\,\big|\,\mathcal{D}^{(j)} \right) = 0, 
    \quad \forall j,
\end{aligned}
\end{equation}
where the canonical EL equation~\eqref{eq:EL-NVT} is imposed as an hard constraint to ensure that $\rho(\bm\theta)$ is strictly belonging to the set of admissible solutions of the augmented cDFT equations in the canonical ensemble. 
The computation of the gradients $\nabla_{\bm \theta}J$ is complicated by the fact that $\rho(\bm\theta)$ is given by $h_c=0$ implicitly, and thus $\nabla_{\bm \theta}\rho$ are not directly available. The optimal parameters $\bm{\theta}^\star$ are thus obtained through a continuous adjoint optimization~\cite{adjoint}. Adjoint optimization allows for arbitrary objective functions, enforcing the EL equation at each training iteration (details in Appendix~\ref{appendix:adjoint_training}). 
\begin{figure}[H]
    \centering
    \includegraphics[width=0.95\textwidth
    ]{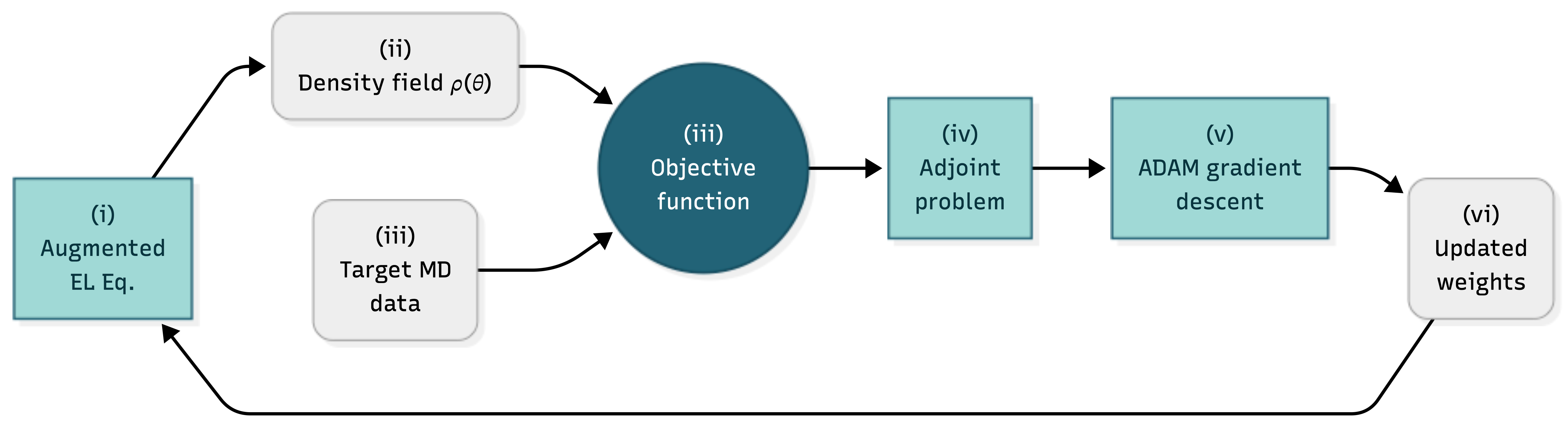}
    \caption{
    Schematic of the adjoint training loop for Eq.~\eqref{eq:Optimization}. At each optimization step, (i) the EL equation~\eqref{eq:EL-NVT} is solved and (ii) the equilibrium density field $\rho(\bm{\theta})$ is computed. Then (iii) the target MD data is injected, and the objective function $J$ in Eq.~\eqref{eq:J} is evaluated. Then (iv) the gradient $\nabla_{\bm{\theta}}J$ is computed in $O(1)$ using adjoint equations ~\eqref{eq:AdjointEquation}. Finally, (v) the ADAM scheme is used to (vi) update the parameters $\theta$.
    }
    \label{fig:online}
\end{figure}
A schematic representation of the training loop is shown in Figure~\ref{fig:online}. It expresses the  "physics-aware" calibration process that preserves the numerical stability of the augmented EL equation~\cite{sirignano}.  
The algorithmic workflow is summarized in Algorithm~\ref{alg:training}. We use ADAM for the updates ~\cite{adam}. Here $\alpha_{\mathrm{LR}}=10^{-3}$ denotes the initial learning rate. We use adaptive schedule, which reduces $\alpha_{\mathrm{LR}}$ when the relative loss $\epsilon_{\mathrm{rel}}=J/J_0$ (with $J_0$ being the objective functional evaluated on the non-augmented density profile $\rho_0$) falls below a prescribed threshold of $\epsilon_\text{T}$.
\begin{algorithm}[H]
\caption{Adjoint-based training loop for cDFT.}
\label{alg:training}
\begin{algorithmic}[1]
    \STATE Initialize $\bm{\theta}$ and compute uncorrected density $\rho_0$.
    \FOR{\textit{each iteration}}
        \STATE Solve DFT EL equations for $\rho(\bm{x};\bm{\theta})$.
        \IF{residuals $<\text{tolerance}$}
            \STATE Solve adjoint equations for adjoint variables $\bm{\lambda}$.
            \STATE Compute gradient $\nabla_{\bm{\theta}} J$.
            \STATE Update parameters via ADAM optimizer.
        \ENDIF
        \STATE Compute relative loss $\epsilon_{\mathrm{rel}} = J/J_0$.
        \IF{$\epsilon_{\mathrm{rel}} < \epsilon_T$}
            \STATE $\alpha_{\mathrm{LR}} \leftarrow \beta_1 \alpha_{\mathrm{LR}}$
            \STATE $\epsilon_T \leftarrow \beta_2 \epsilon_T$
        \ENDIF
    \ENDFOR
    \STATE Save optimized parameters $\bm{\theta}^\star$.
\end{algorithmic}
\end{algorithm}

\subsection{Results: Improved Adsorption Density Profiles, Bulk Properties, and Surface Tension}
\label{subsec:calibration_results}

The results obtained using the calibrated $\bar F_\text{ex}$ are presented below.  
The bulk properties and inhomogeneous features are assessed against MD reference data $\mathcal D_\text{ref}$ and the baseline MF formulation. Unless otherwise stated, all reported quantities are given in LJ reduced units.

\subsubsection{Adsorption Density Profiles}
\begin{figure}[hbt!]
    \centering
    \includegraphics[width=0.75\textwidth]{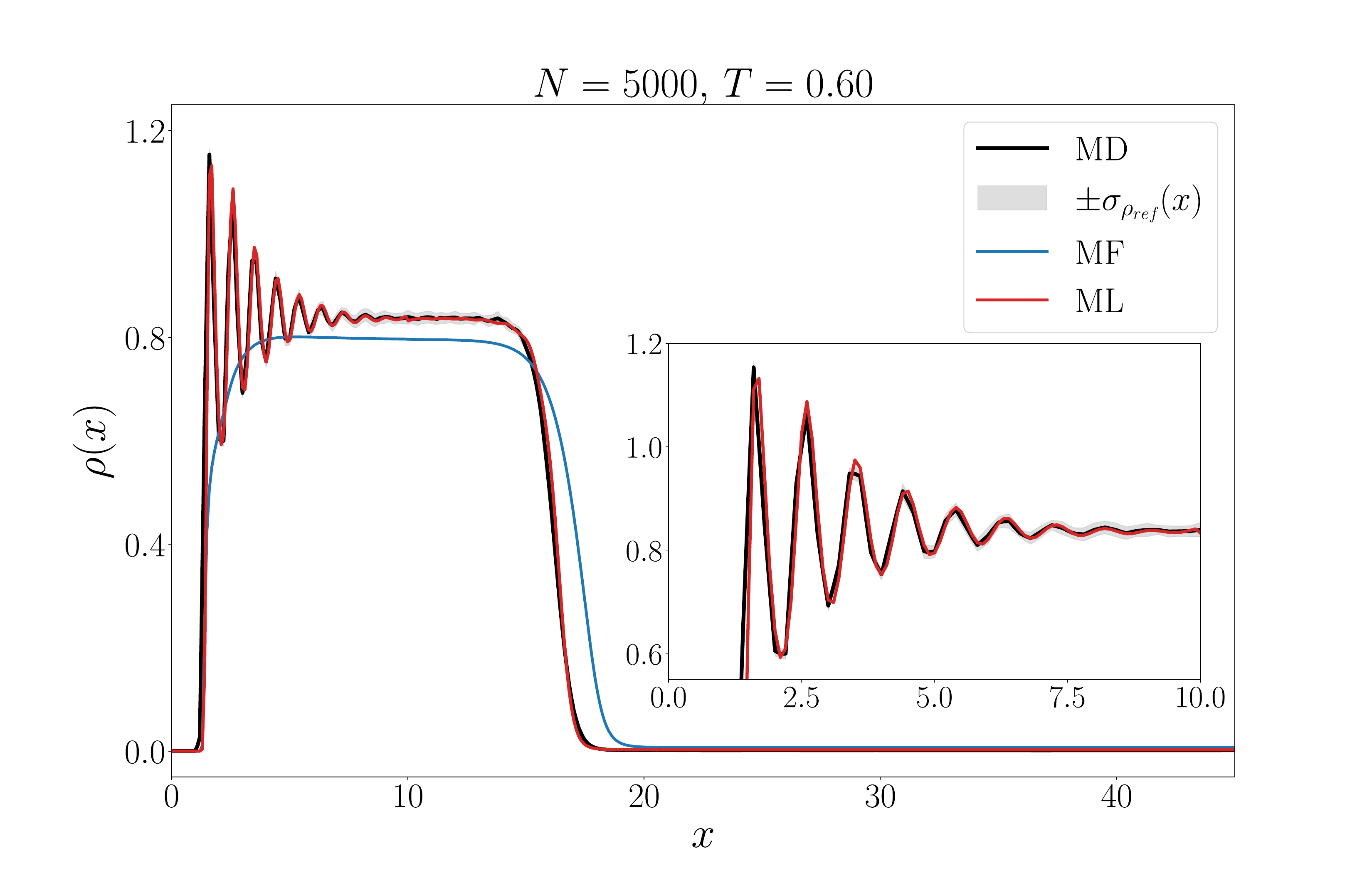}
    \caption{
        \textcolor{black}{
    Superposition of wall-normal equilibrium density profiles for validation data at $N=5000,\,T=0.60$: MD reference (black, shaded confidence interval), baseline mean-field cDFT (blue), and ML-augmented cDFT (red).
    The validation MD profile $\rho_\text{ref}(x)$ is shown with shaded confidence bands $\rho_\text{ref}(x)\pm  \sigma_{\rho_\text{ref}}(x)$, where $\sigma_{\rho_\text{ref}}(x)$ is the standard deviation at position $x$.
    The ML cDFT accurately captures the near-wall layering and the bulk densities at the plateaus of $\rho_\text{ref}(x)$. On the other hand, the baseline DFT misses all these features.  
    }}
    \label{fig:rho_trained}
\end{figure}

Figure~\ref{fig:rho_trained} superimposes a wall-normal density profile $\rho_\text{ref}\in\mathcal{D}_\text{ref}$ from the MD validation dataset with predictions from the baseline MF cDFT $\rho_\text{mf}$, and the ML-augmented cDFT $\rho$.
The profiles are representative of our experiments. 
The MD reference profile is reported together with its statistical confidence band $\rho_\text{ref}(x)\pm\sigma_{\rho_\text{ref}}(x)$, where $\sigma_{\rho_\text{ref}}(x)$ is the standard deviation at position $x$.
The results shown are taken outside of the training data set and demonstrate a remarkable agreement between the MD and ML cDFT. Especially striking is the fact that a single WDA-like kernel enables our functional to capture the intricate details of the near-wall fluid structure after being appropriately calibrated. 
The baseline MF cDFT captures the qualitative features of adsorption, but obviously cannot recover the near-wall oscillations as well as the bulk density values. 

\begin{figure}[hbt!]
    \centering
    \includegraphics[width=0.9\textwidth]{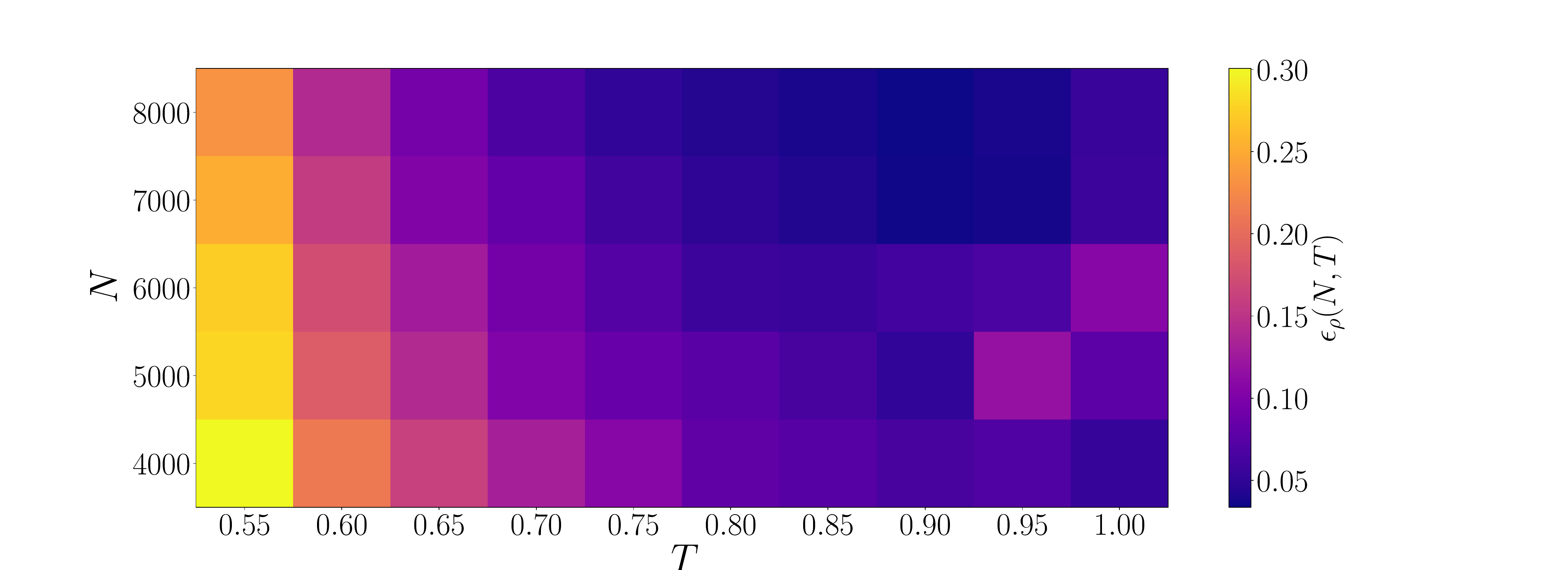}
    \caption{
    Heat map of the relative error $\epsilon_\rho(N,T)$~\eqref{eq:eps_rho} across the validation set. The color scale reports the deviation between the augmented cDFT predictions and MD reference density profiles as a function of particle number $N$ and temperature $T$.
    In particular, the trained model is able to reduce the $L_1$ error between the ML prediction and MD of around $90\%$ with respect to the for a wide range of the testing data set. 
    }
    \label{fig:err_rel}
\end{figure}
\noindent
To assess accuracy across the full validation set $\mathcal D_t$, Figure~\ref{fig:err_rel} shows the relative error in the $L_1$ norm:
\begin{equation}
\epsilon_\rho(T,N)=
\frac{\|\rho(\cdot;T,N)-\rho_\text{ref}(\cdot;T,N)\|_{L_1}}
{{\|\rho_{mf}(\cdot;T,N)-\rho_\text{ref}(\cdot;T,N)\|_{L_1}}},
\label{eq:eps_rho}
\end{equation}
which quantifies the improvements with respect to the baseline MF model.
Across the entire validation dataset, the ML-augmented cDFT yields $\epsilon_\rho < 1$, demonstrating a consistent improvement over the baseline model. The reduction in error is more pronounced at high temperatures, where the baseline model is less accurate and the improvements is more marked. However, the ML-corrected functional
remains significantly more accurate than MF cDFT throughout, reducing the baseline error of around $90\%$ for a wide range of the testing data set.

\subsubsection{Liquid-Vapor Bulk Coexistence}
\begin{figure}[hbt!]
    \centering
    \includegraphics[width=0.55\textwidth]{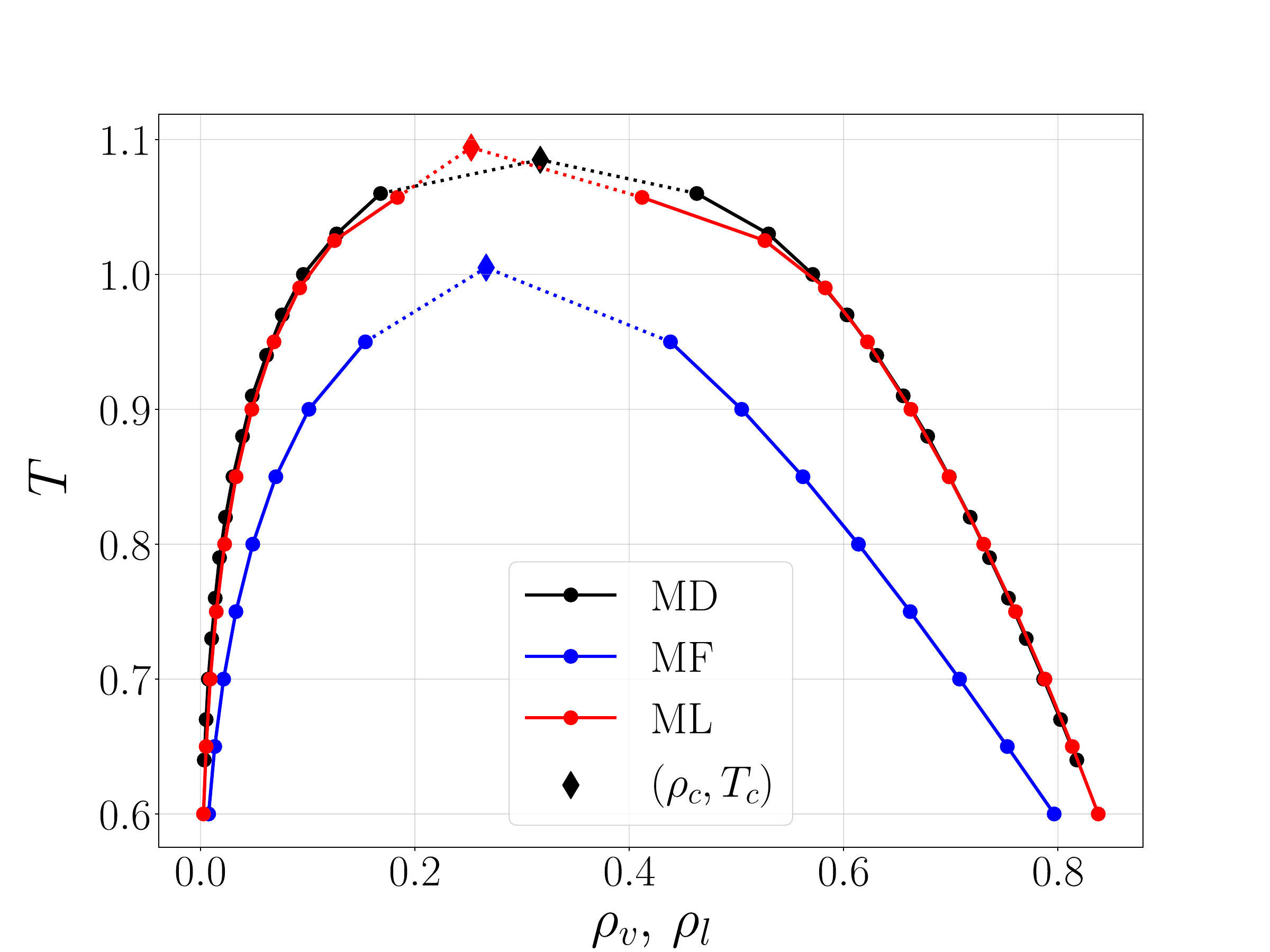}
    \caption{
    Superposition of bulk liquid-vapor binodals: MD (black circles),  baseline MF cDFT (blue circles), and ML cDFT (red circles). Our ML cDFT closely reproduce the MD binodal, and substantially improves upon the systematic underestimation of the MF model, recovering the temperature of the critical point (diamonds) within approximately $1.5\%$ relative error. 
    This is remarkable, because the model was trained only using subcritical temperatures $T\leq0.95$. 
    }
    \label{fig:coex}
\end{figure}

To compute the bulk binodal for our model fluid, we solve the coexistence equations, equating the pressures $p$ and chemical potentials $\mu$ in the coexisting phases:
\begin{equation}
\label{eq:coex_conditions}
\begin{cases}
\mu(\rho_l, T ) &= \mu(\rho_v, T),\\
p(\rho_l, T) &= p(\rho_v, T),
\end{cases}
\end{equation}
where $\rho_l$ and $\rho_v$ are the densities of coexisting liquid and gas.
We trained ML cDFT on adsorption density profiles, taken at five different $T$: $0.55$, $0.65$, $0.75$, $0.85$ and $0.95$. Nevertheless, our model shows excellent generalization in predicting bulk coexistence densities across temperature. Figure~\ref{fig:coex} shows the liquid–vapor coexistence curve $(\rho_l(T),\rho_v(T))$ obtained from Eqs~\eqref{eq:coex_conditions}. For comparison, we superimpose the tabulated LJ binodal, carefully obtained from specialized MD simulations ~\cite{comprehensive_study_of_the_vapour–liquid_coexistence}. Our model shows remarkable agreement with the results in \cite{comprehensive_study_of_the_vapour–liquid_coexistence}. 

Particularly surprising is the fact that we are able to recover the critical temperature, extrapolating outside the training temperature range. The critical point $(T_c, \,\rho_c)$ satisfies the following condition:
\begin{equation}
\label{eq:critical condition}
\left[\cfrac{\partial p}{\partial \rho}\right]_{T=T_c} = \left[\cfrac{\partial^2 p}{\partial \rho^2}\right]_{T=T_c}= 0.
\end{equation}

We obtain $(T_c= 1.094, \,\rho_c= 0.2523)$. The value of $T_c$ is in agreement with the values established in detailed simulation studies, e.g.  Smit~\cite{Smit} and Vrabec et al.~\cite{comprehensive_study_of_the_vapour–liquid_coexistence} report $(T_c=1.085, \,\rho_c=0.3190)$ and $(T_c=1.0779, \,\rho_c=0.317)$, measuring a relative error of $\epsilon_{c,1}= 0.83\%$ and $\epsilon_{c,2}= 1.34\%$ respectively. 
On the other hand, the baseline mean-field model report $(T_c= 1.005, \,\rho_c= 0.2665)$, strongly underestimating the critical temperature.

\subsubsection{Liquid-Vapor Surface Tension}
\label{subsec:surface_tension_lg}
\begin{figure}[h!]
    \centering
    \includegraphics[width=0.55\textwidth]{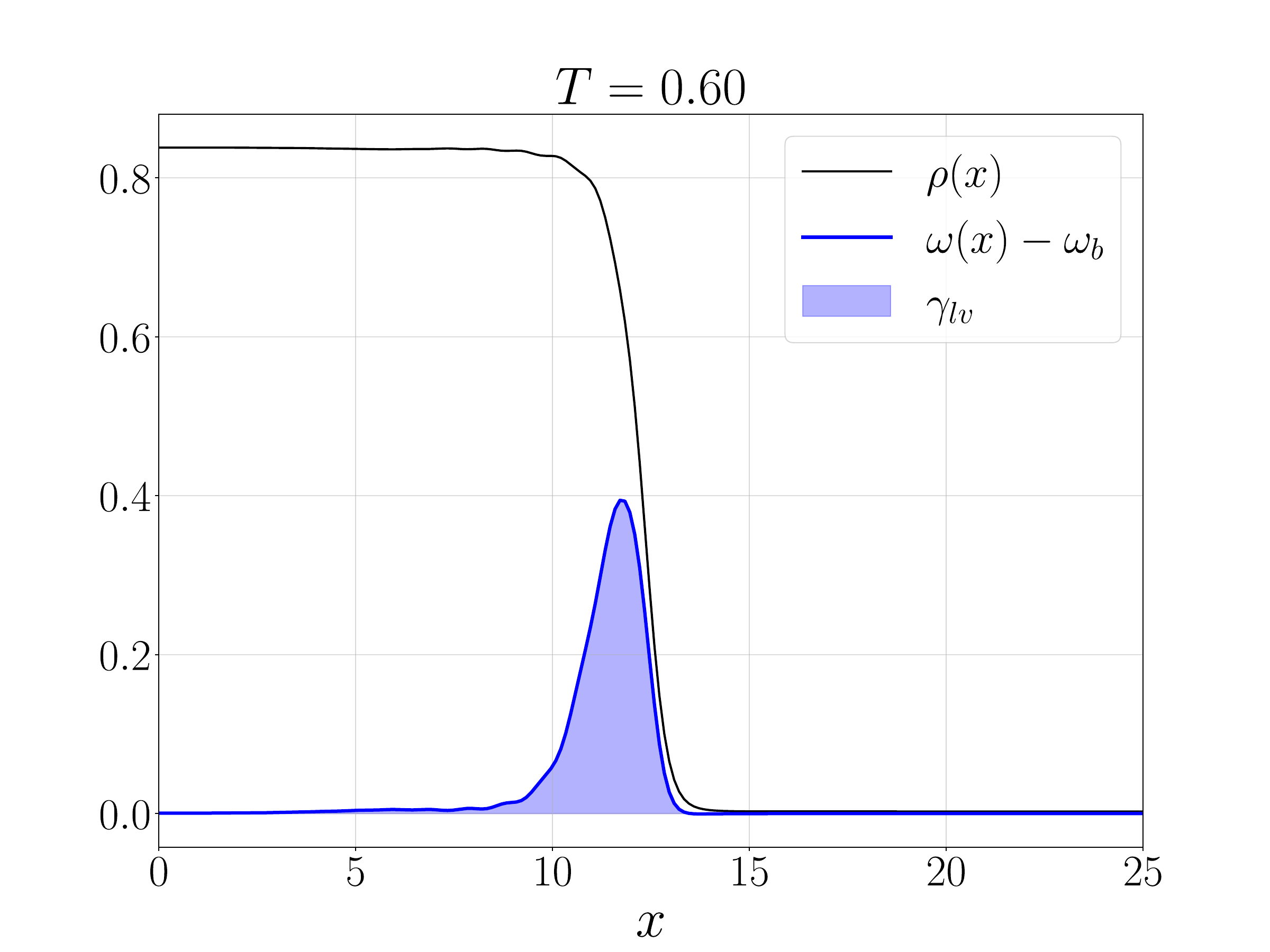}
    \caption{
    Planar profiles of fluid density and excess free energy density at bulk liquid-vapor coexistence. The blue area gives the value of the liquid-gas surface tension. 
    }
    \label{fig:omega_ex_gamma}
\end{figure}

We can obtain the density profile of coexisting liquid and vapor by solving the EL equation~\eqref{eq:EL-NVT} at coexistence, subject to the boundary conditions of contact with liquid and vapor on the opposite sides of the computational domain. The liquid-vapor surface tension is then given by integrating the excess over the bulk of the grand free energy density $\omega$ (see Figure~\ref{fig:omega_ex_gamma}):
\begin{equation}
\label{eq:gammalg}
    \gamma_{lv} = \int_{-\infty}^{+\infty} \big( \omega[\rho] - \omega_b \big) \, dx.
\end{equation}
where $\omega_b$ denotes the bulk grand free-energy density at coexistence, which is identical in the liquid and vapor phases and may be evaluated at either $\rho_l$ or $\rho_v$.
 
\begin{figure}[h!]
    \centering
    \includegraphics[width=0.55\textwidth]{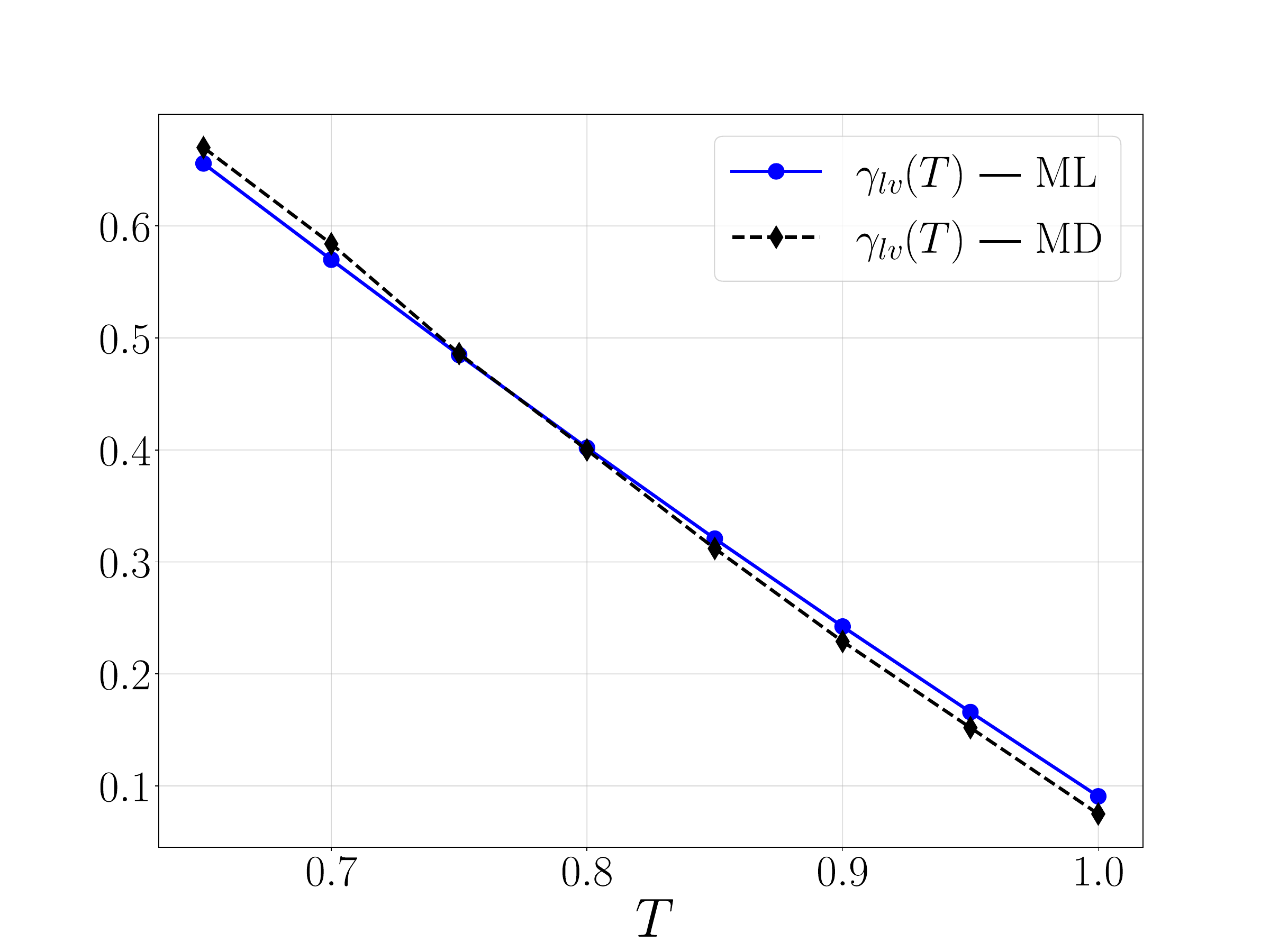}
    \caption{
   {Liquid-vapor surface tension $\gamma_{lv}(T)$ computed from ML cDFT (blue), superimposed against published MD data in \cite{lj_gamma} (black). 
    }}
    \label{fig:gamma_vs_T}
\end{figure}

Figure~\ref{fig:gamma_vs_T} shows the comparison of the temperature dependence of the surface tension obtained from ML cDFT with the MD simulation results of Stefan et al.~\cite{lj_gamma}. We see a remarkable agreement across a range of temperatures. 
Notably, our model managed to capture the free liquid-gas interface and its surface tension by being trained only on adsorption data. 
 
%



\section{Upscaling to 3D Droplets Shape Prediction}
\label{subsec:droplets}
In what follows, we apply our theory to investigate adsorbed liquid drops. We demonstrate the remarkable ability of our simple $\bar F_\text{ex}$ model to capture the apparent contact angles $\theta_c$ of nano-drops and the wetting properties of the substrate. To do so, we develop a Laplace-like description of adsorption, where the microscopic information is subsumed into the interface binding potential~\cite{Dietrich, Yatsyshin2018, nanodrop_peter}.

\subsection{From Planar Adsorption to the 3D Droplet Free Energy}
\begin{figure}[h!]
    \centering
    \subfloat[Non-interacting interfaces: $\Delta\gamma(l)\approx 0$]{\includegraphics[width=0.47\textwidth]{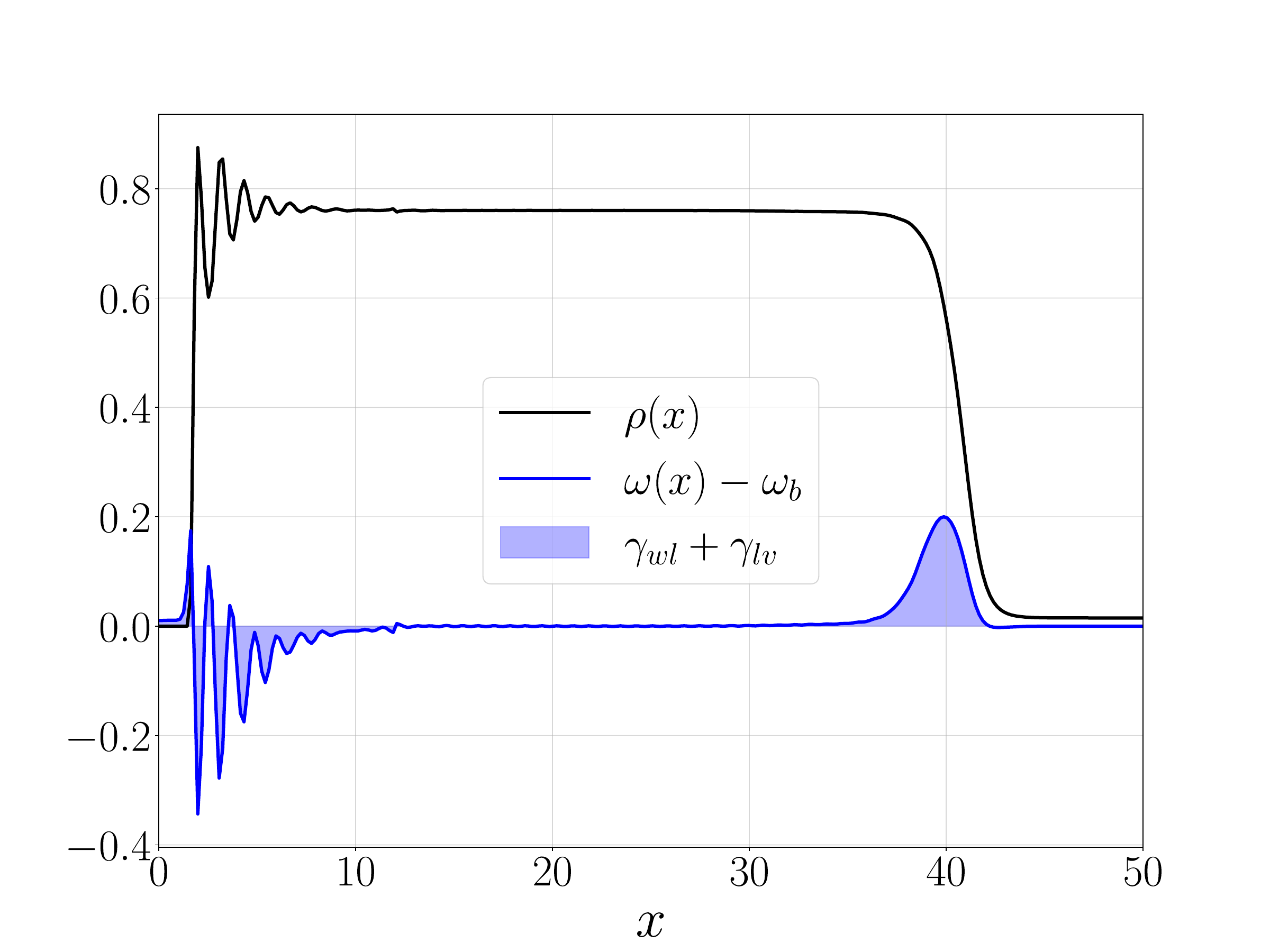}}
    \hspace{0.05\textwidth}
    \subfloat[Interacting interfaces: $\Delta\gamma(l)\neq 0$]{\includegraphics[width=0.47\textwidth]{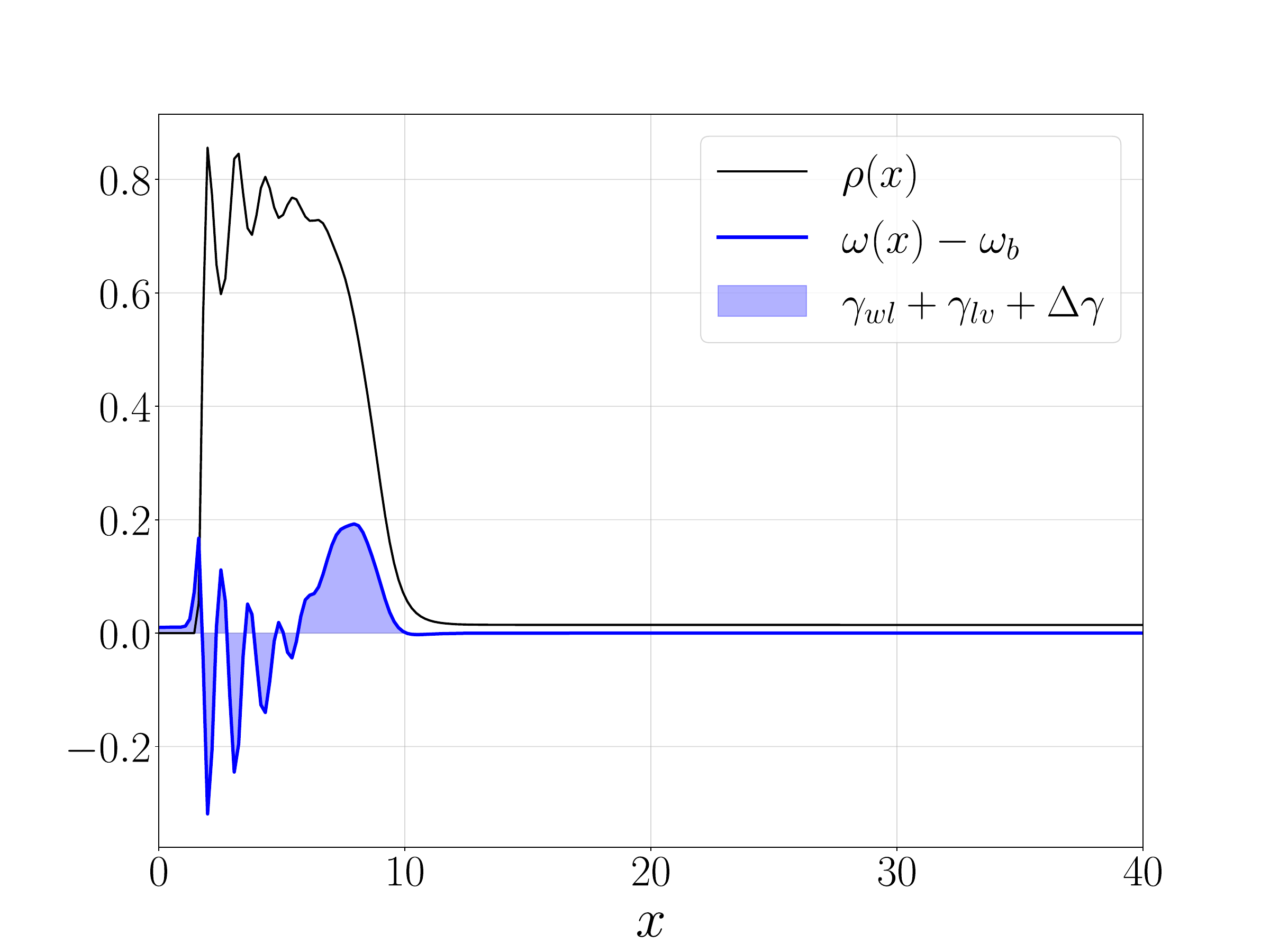}}
    \caption{   
    $W_0(l)$ is given by the area under the graph of $\omega(x) - \omega_b$.   
    (a) Thick adsorbed film. The wall–liquid and liquid–gas interfaces do not interact, leading to $W_0=\gamma_{wl} + \gamma_{lv}$.  
    (b) Thin adsorbed film. The interaction of the wall-liquid and liquid-vapor interface is captured by introducing the binding potential $\Delta\gamma(l)$, leading to Eq.~\eqref{eq:Wgam}.}
    \label{fig:W_decomposition}
\end{figure}

\begin{figure}[hbt!]
    \centering
    \includegraphics[width=0.55\textwidth]{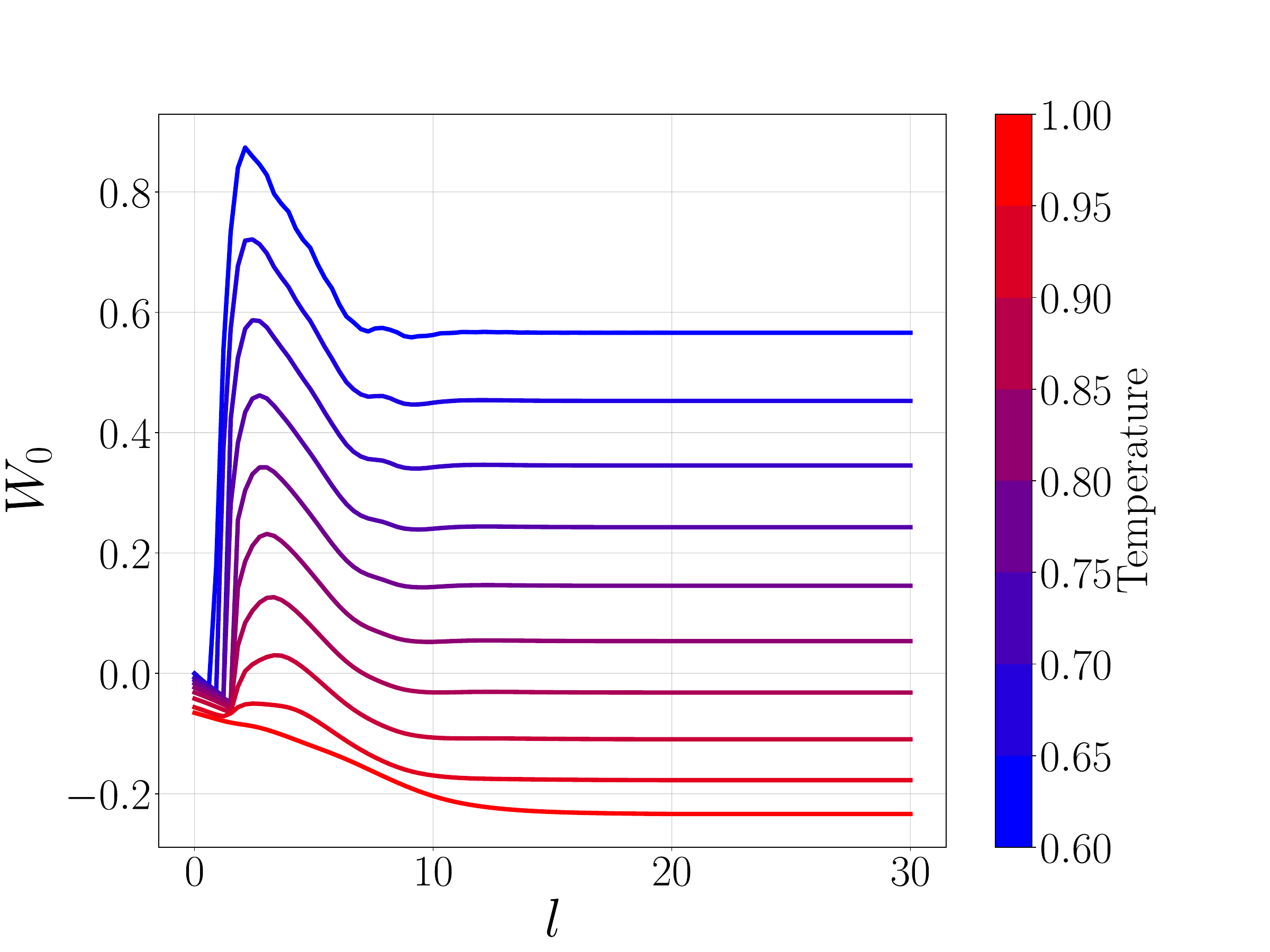}
    \caption{
    Work of adsorption in Eq.~\eqref{eq:W0}, computed for a range of temperatures using $\bar F_\text{ex}$. The position of the dominant minimum determines the height of the adsorbed film. At higher temperatures, the dominant minimum shifts to larger $l$, signaling a transition to wetting.}
    \label{fig:W_T_dep}
\end{figure}

In an isothermal setting, the family of equilibrium planar density profiles $\rho_0(x;\mu, T)$ obtained by varying the chemical potential $\mu$, can be obtained using a continuation algorithm over $\mu$. To describe isothermal adsorption, the film height $l$ is used as an order parameter for film growth, with the chemical potential $\mu$ serving as the control parameter. The thermodynamic work of adsorbate formation is given by the integral of the excess-over-bulk grand potential density:
\begin{equation}
\label{eq:W0}
    W_0(l;\;T) = \int_{0}^{+\infty} 
    \left[\,(\omega\big([\rho_0(\,\cdot\,;\;l)];\;T,\mu_0(l)\big) - \omega\big([\rho_b
    ];\;T,\mu_0(l)\big))\,\right] \, dx ,
\end{equation}
where $\rho_b$ is the bulk gas density. 
Given the adsorption density profile $\rho$, the film height can be obtained from the following equation:
\begin{equation}
   \label{eq:l}
   l = \frac{\int_0^{+\infty} \big(\rho(x) - \rho_b\big)\,dx}{\rho_l - \rho_v},
\end{equation}
where $\Gamma = \int_0^\infty \big(\rho(x) - \rho_b\big)\,dx$ is adsorption. 
The macroscopic point of view is to treat interfaces as sharp, casting $W_0(l, T)$ in terms of surface tensions:
\begin{equation}
\label{eq:Wgam}
    W_0(l;T) = \gamma_{wl}(T) + \gamma_{lv}(T) + \Delta\gamma(l,T),
\end{equation}
where $\gamma_{wl}$ and $\gamma_{lv}$ denote the wall–liquid and liquid–vapor surface tensions, respectively, and $\Delta\gamma(l,T)$ is the so-called interface binding potential that accounts for non-locality of molecular interactions, as the interaction between wall-liquid and liquid-vapor interfaces. This is illustrated in Figure~\ref{fig:W_decomposition} showing the density profiles with large and small adsorptions. As $\Delta\gamma(l)\!\to\!0$ with $l\to\infty$, $W_0$ reduces to the sum of surface contributions $\gamma_{wl}+\gamma_{lv}$. 
Figure~\ref{fig:W_T_dep} represents a number of potential surfaces $W_0(l)$, computed using $\bar F_\text{ex}$, at several different $T$. Notice that as temperature increases, the minima of $W_0(l)$ become shallower and shift towards larger $l$. In the limit, the dominant minimum of $W_0(l)$ occurs at $l\to\infty$, signifying transition to wetting. To obtain the total work of adsorption at a given chemical potential $\mu$ and temperature $T$, we need to add the bulk contribution to $W_0$:
\begin{equation}
     W(l;T,\mu) = W_0(l;T) - (\mu - \mu_0(l))(\rho_l-\rho_v)\, l.
     \label{eq:W_mu}
\end{equation}
In practice, computing $W_0(l)$ allows us to obtain the wall-fluid surface tensions as well. The asymptotic limit of $W_0(l)$ provides $\gamma_{wl}(T)$:
\begin{equation}
    \gamma_{wl}(T) = \lim_{l\to\infty} W_0(l;T) - \gamma_{lv}(T),
    \label{eq:gamma_wl_short}
\end{equation}
The wall–vapour surface tension $\gamma_{wv}$ is obtained from Eq.~\eqref{eq:W0} by using the density profile of the coexisting vapor in contact with the wall,  $\rho_{wv}(z)$, and is approximately equal to $W_0(l=0, T)$:
\begin{equation}
    \gamma_{wg}(T) = \int_{-\infty}^{+\infty }
    [\,\omega(\rho_{wv}(z);T,\mu_0) - \omega(\rho_v;T,\mu_0)\,]\,dz .
    \label{eq:gamma_wg_short}
\end{equation}
The temperature dependence of surface tensions, obtained from ML cDFT, is represented in Figure~\ref{fig:gamma_vs_T}. The surface tensions exhibits decay with temperature. The value of $T_w$, at which $\gamma_{wl}(T_w)+\gamma_{lv}(T_w)-\gamma_{wv}(T_w)=0$ identifies the limiting wall wetting temperature $T_w=0.8534$. 
\begin{figure}[hbt!]
    \centering
    \includegraphics[width=0.55\textwidth]{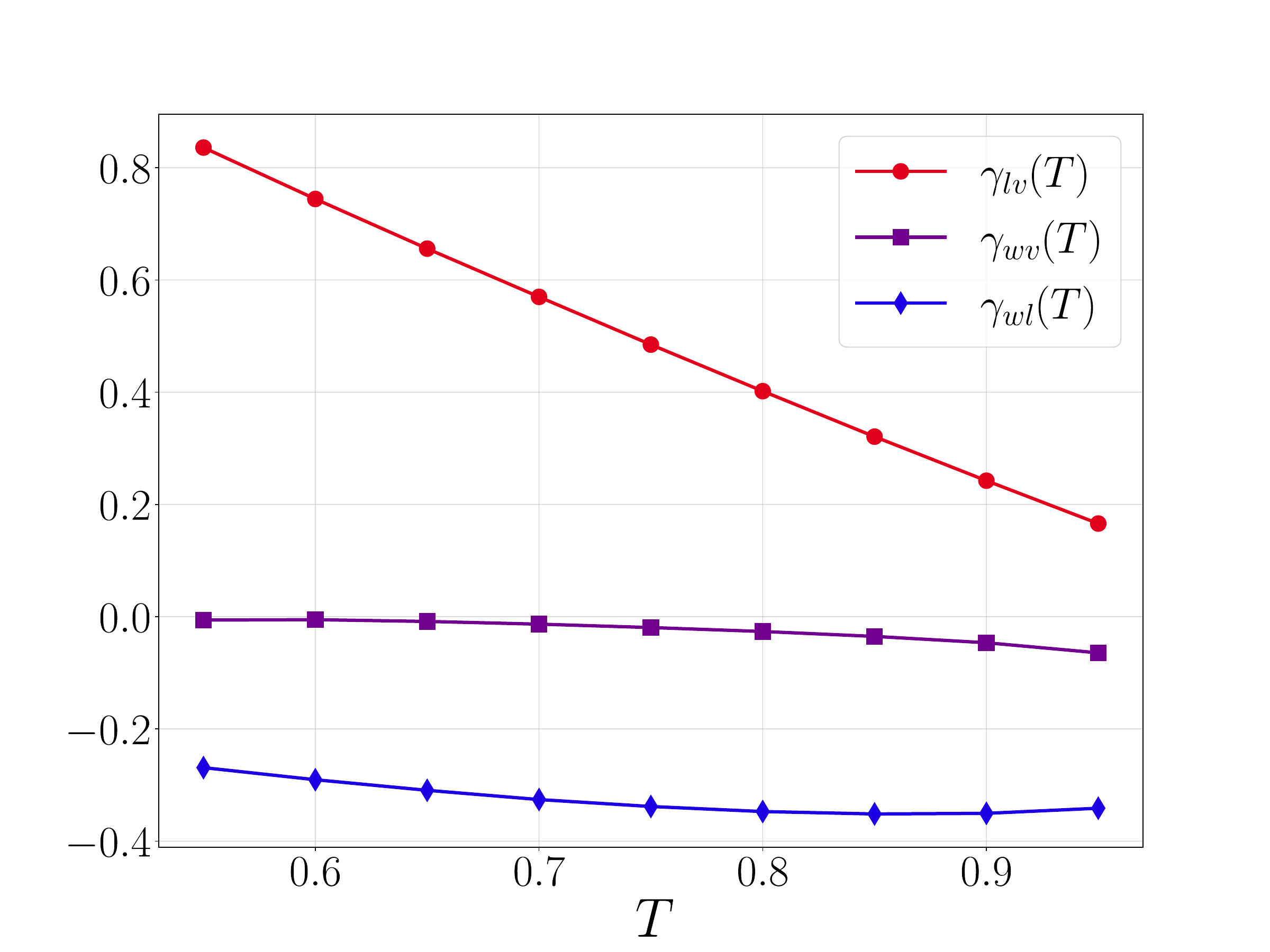}
    \caption{
    Temperature dependence of the interfacial surface tensions.  
    Solid lines: $\gamma_{lv}(T)$ (red), $\gamma_{wv}(T)$ 
    (purple), and $\gamma_{wl}(T)$ (blue).  
    The condition $\gamma_{wl}+\gamma_{lv} - \gamma_{wv}=0$ identifies the wall wetting temperature $T_w=0.8534$.}
    \label{fig:gammas_vs_T}
\end{figure}

\subsubsection{Contact Angles of Liquid Drops}

In the sharp-interface limit, the excess free energy of adsorbed drop is given by the following expression:
\begin{equation}
    \Delta \Omega(T,\mu) = 
    \gamma_{lv} S_c + (\gamma_{wl}-\gamma_{wv})A_c 
    + \int_{A_c} \!\!\Big[\Delta\gamma(l)
    - (\mu - \mu_0)(\rho_l-\rho_v)\,l \Big]\! dA ,
    \label{eq:omega_droplet}
\end{equation}
where $S_c$ and $A_c$ are the liquid–vapor and wall-liquid contact areas, respectively. The integral term contains the contribution of molecular-level interactions, such as line tension, which are not typically included in macroscopic descriptions based on Young-Laplace equation~\cite{laplace}. In practice, Eq.~\eqref{eq:omega_droplet} locally approximates adsorbate as a composition of planar wetting slabs of volume $ldA$, leading to surprisingly accurate models of adsorbed nano-drops and nano-bubbles~\cite{nanodrop_peter}.

\begin{figure}[h!]
    \centering
    \subfloat[]{\includegraphics[width=0.4\textwidth, clip]{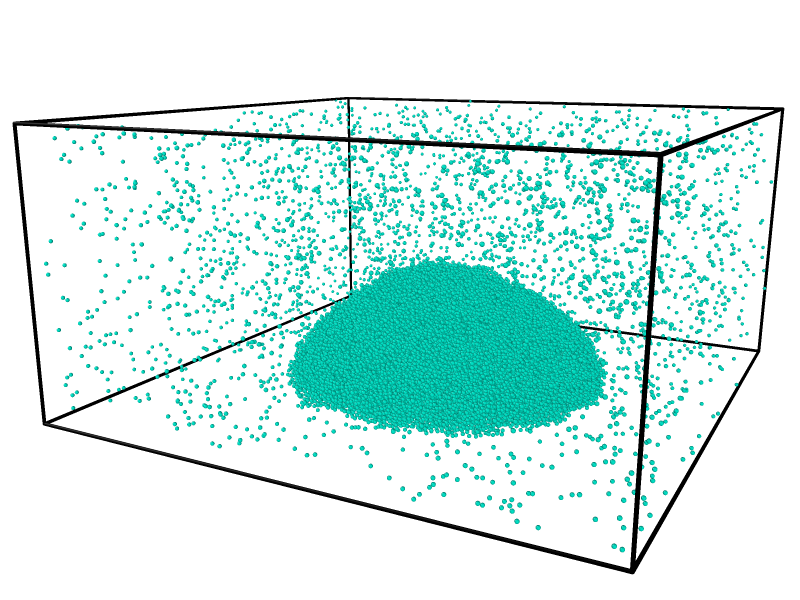}}
    \hspace{0.05\textwidth}
    \subfloat[]{\includegraphics[width=0.4\textwidth, clip]{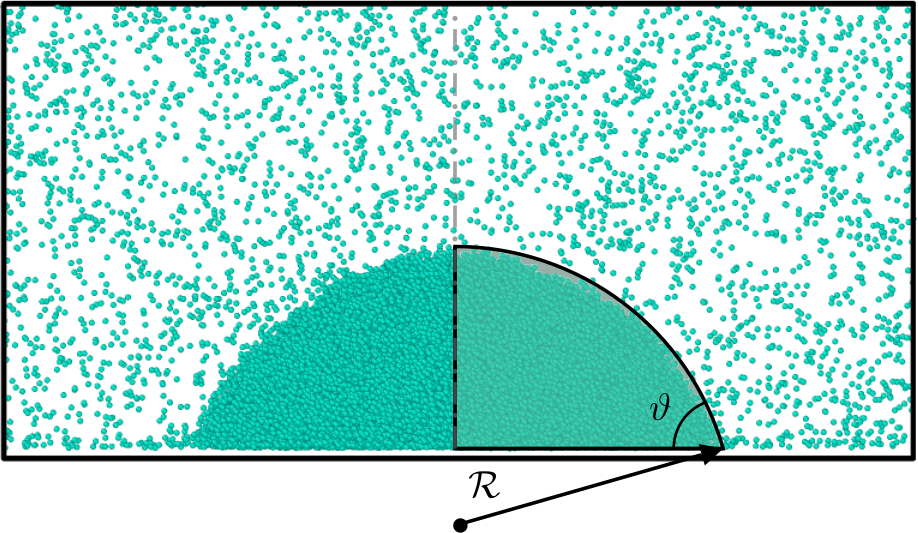}}
    \caption{Illustrative example of MD data showing a liquid droplet on a substrate, in contact with its saturated vapor: (a) perspective view, (b) lateral view, highlighting the droplet spherical cap parametrization in terms of the radius $\mathcal R$ and the contact angle $\vartheta$. 
    The simulation was initialized as a spherical cap resting on the wall. An equilibration run is performed to relax the system to its equilibrium state. After equilibrium is reached, a production run is carried out to collect the statistical data for analysis.}
    \label{fig:md_droplet_box}
\end{figure}

Assuming the droplet has a spherical cap, its interface $\mathcal{S}_\text{cap}$ can be parametrized by the cap radius $\mathcal R$ and the contact angle $\vartheta$, as illustrated in Figure~\ref{fig:md_droplet_box}. We fix the number of particles $N$ to directly compare to MD simulations. The equilibrium configuration $(R,\theta_c)$ minimizes the excess-over-the-bulk HFE $\Delta F(\mathcal R,\vartheta) =\Delta\Omega(\mathcal R,\vartheta) + \mu \Delta N$, where $\Delta N=N-\rho_v V$ is the number of excess particles that form the liquid drop inside the simulation box of volume $V$. The equilibrium radius $R$ and apparent angle $\theta_c$ of the equilibrium drop are obtained from the following minimization problem, posed in the $NVT$ ensemble:
\begin{equation}
\begin{aligned}
      (R,\theta_c) &=\argmin_{\mathcal R,\vartheta} \Delta F(\mathcal R,\vartheta),\\
\text{subject to}\quad &(\rho_l- \rho_v)V_{drop}(\mathcal R,\vartheta) - \Delta N = 0.
\end{aligned}
\end{equation}
Minimizing $\Delta F$ provides the mapping between the canonical thermodynamic state and the apparent contact angle $(N,T) \to\ \theta_c(N,T)$.
To compute $\Delta F$, we only require $W_0$~\eqref{eq:W0}, which can be obtained from planar adsorption isotherms. This is highly computationally efficient and allows us to obtain the interfaces of drops in a wide range of  sizes and temperatures.

The resulting function $\theta_c(N,T)$ fully captures the wall wetting behavior at different length scales, from nano- to macroscopic scales. At large $N$, $\theta_c $ approaches the Young contact angle $\theta_Y$ by construction. We validate our approach by the direct comparison with MD simulations.

\subsubsection{Molecular Dynamics Simulation of Nano-Drops}
\label{subsec:MDdroplet}

Our MD simulations were carried out using the LAMMPS package~\cite{LAMMPS}. We simulated our drops in the $NVT$ ensemble with a Langevin thermostat. 
The fluid–fluid interactions were given by the truncated and shifted LJ12--6 potential in Eq.~\eqref{eq:lj126} with $r_c=2.5$. The MD fluid model must match the cDFT one. 
The boundary conditions were reflective at the top of the simulation box and periodic on the sides. The simulation box was rectangular with dimensions $L_x=L_y=120$ and $L_z=60$. This size is sufficiently large to avoid the effects of periodic boundary conditions on the drop shape. The fluid-substrate interaction is given by LJ9--3 potential in Eq.~\eqref{eq:lj93} with parameters $\bm p_w^{(2)}=[ \varepsilon_w=1,\,\sigma_w=2]$.  

An illustrative example of the simulation setup is shown in Fig.~\ref{fig:md_droplet_box}b. The fluid-wall particle potential has to be the same as used in Eq.~\eqref{eq:W0} to compute the planar adsorption isotherms. This ensures that all the microscopic information is appropriately captured by the interface binding potential. However, the external particle potential does need to be the same as the one used to train the cDFT correction terms described in section \ref{sec:DataDrivenHFE}, allowing the model to be highly transferable. The MD density profiles were obtained by binning the atomic trajectories in cylindrical coordinate system and averaging appropriately over time.

\subsection{Shape and Contact Angle Prediction}
\label{subsec:results_drop}

\begin{figure}[h]
    \centering
    \subfloat[$T=0.6$]{\includegraphics[width=0.47\textwidth]{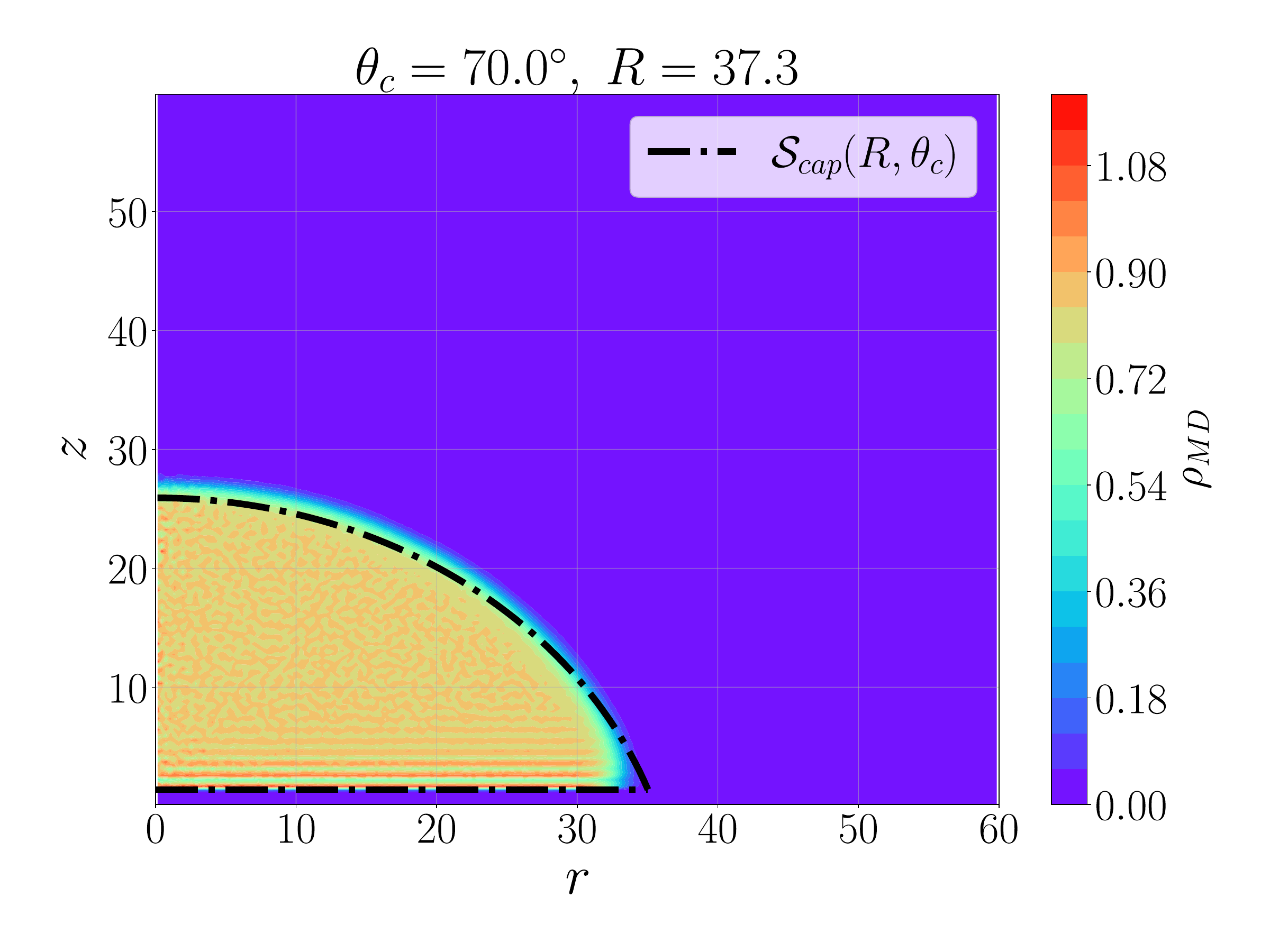}}
    \hspace{0.05\textwidth}
    \subfloat[$T=0.7$]{\includegraphics[width=0.47\textwidth]{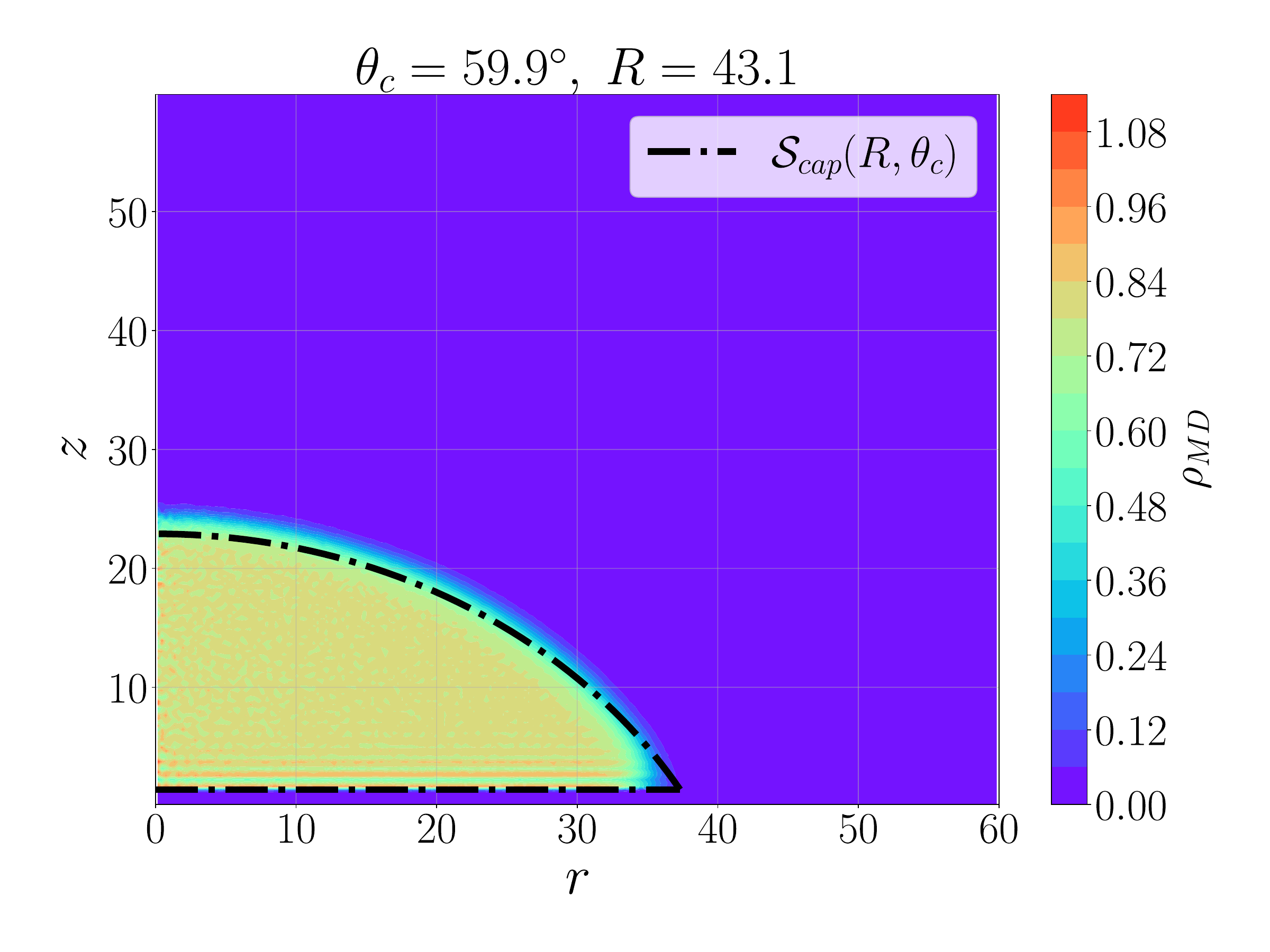}}

    \subfloat[$T=0.8$]{\includegraphics[width=0.47\textwidth]{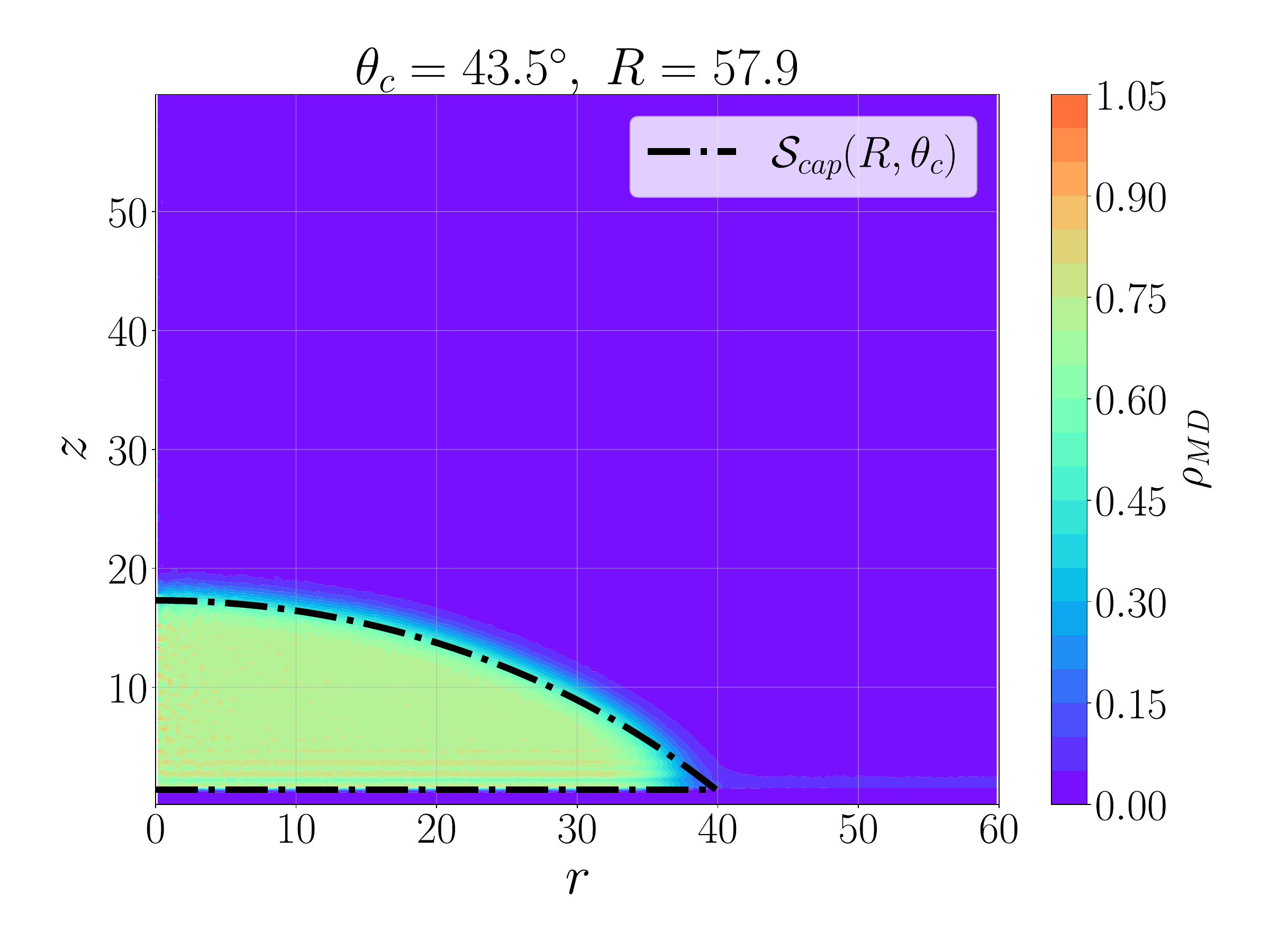}}
    \hspace{0.05\textwidth}
    \subfloat[$T=0.9$]{\includegraphics[width=0.47\textwidth]{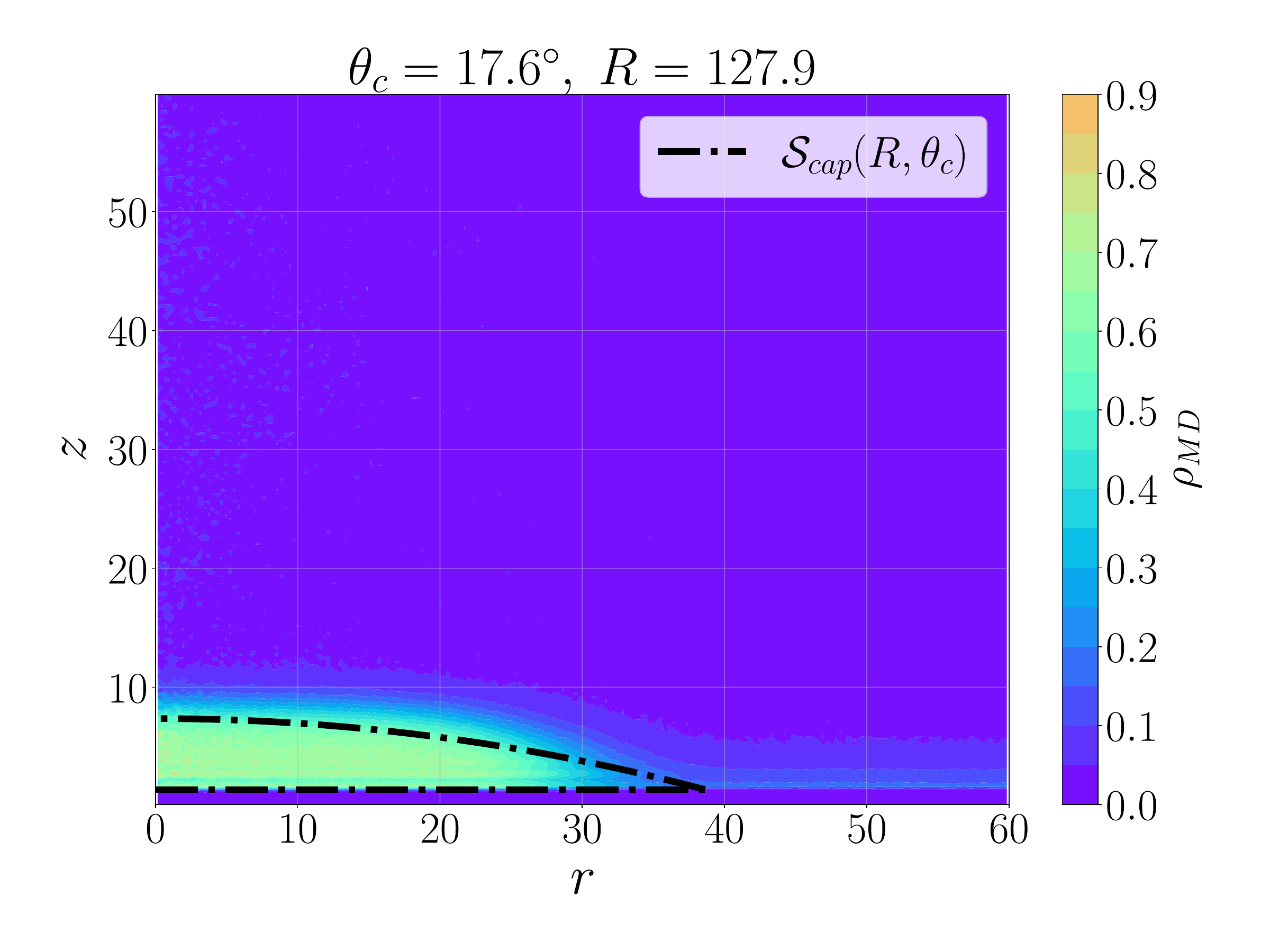}}
    
    \caption{
        Our model, based on effective binding potential, which was in turn obtained from calibrated ML DFT shows good agreement with MD simulation. Panels (a)-(c) show the superposition of droplet shapes  (dashed contour) and density profiles from MD simulations (colored contours). As expected, the computed contact angle decreases with temperature.
    }
    \label{fig:droplet_results}
\end{figure}

Our methodology reduces the need for MD in studies targeting contact lines across droplet sizes, and offers principled means to bridge microscopic treatments of adsorption with continuum-mechanical approaches, such as Navier-Stokes. In Figure~\ref{fig:droplet_results} we visually assess the performance of our binding potential model by superimposing the equilibrium spherical drop cap, determined by the $\theta_c$ and $R$, over the MD simulation the same drop. We observe good agreement of our theoretical model with simulation: the liquid-gas and wall-gas interface is captured, and the apparent contact angle of the nano-drops agrees with the simulations. Figure~\ref{fig:droplet_results} represents this comparison for a range of temperatures and drop sizes. 

We find good agreement across a wide range of thermodynamic states using a wall potential not belonging to the training set, demonstrating that our model is capable of generalize and captures both microscopic and macroscopic wetting features.
\begin{figure}[h]
    \centering
    \includegraphics[width=0.55\textwidth]{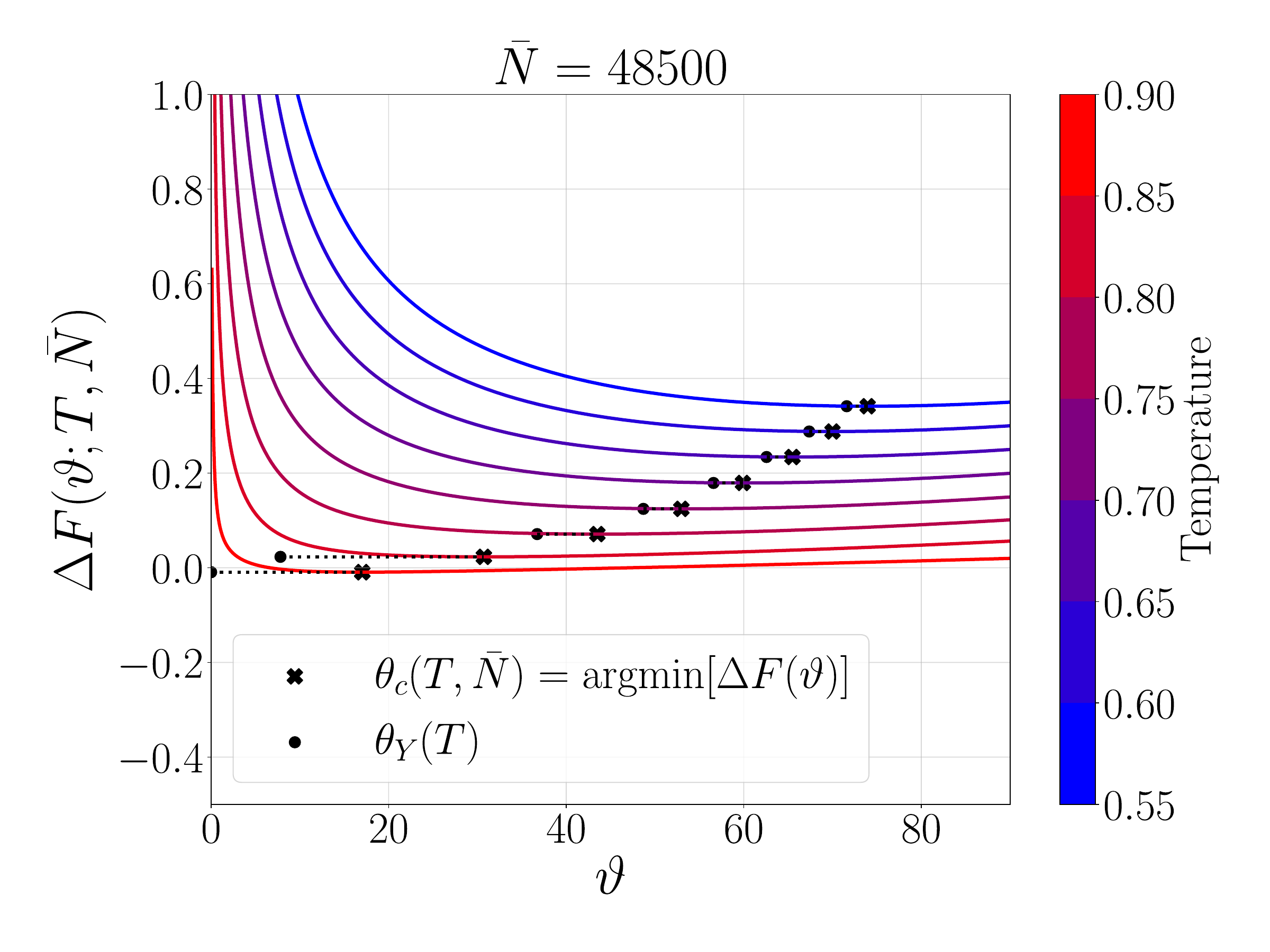}
    \caption{\textcolor{black}{
    Constant-$T$ level sets of $\Delta F(\vartheta;T,\bar N)$-surface, evaluated at $\bar N = 48500$ and normalized by the wall area. Minima of each level set identify the apparent contact angle of nanodrops (crosses), which are different from the macroscopic Young angles $\theta_Y$ (dots). Wetting transition happens near $T=0.8534$, where the Young contact angle $\theta_Y$ tends to $0$.   
    }}
    \label{fig:DF_vs_theta}
\end{figure}
However, the deviation from MD at high $T$ are present, as visible in Figure~\ref{fig:droplet_results}d. Near the drop contact line, where the liquid-vapor interface meets the wall, the binding potential affects the interface curvature and thereby violates the spherical-cap assumption. This effect may be seen in all examples represented in Fig.\ref{fig:droplet_results}, but are most pronounced at high $T$, where the drop height is low. 

Figure~\ref{fig:DF_vs_theta} shows the constant-$T$ contours of $\Delta F$ as functions of the parametrized apparent contact angle $\vartheta$ at $\bar N=48500$.  The minimum of each constant-$T$ level set, which correspond to the equilibrium apparent contact angle $\theta_c$, is designated with a cross. And we also show the respective Young angle $\theta_Y$.
When $T\to T_w$, the global minima of the constant-$T$ level sets tend to $\theta_c=0^\circ$ as the adsorption, $\Gamma$, exhibits a discontinuous singularity. 
It is also noteworthy that at low temperatures, the apparent angle $\theta_c$ tends to its macroscopic Young counterpart $\theta_Y$. 

\begin{figure}[h!]
    \centering
    \includegraphics[width=0.55\textwidth]{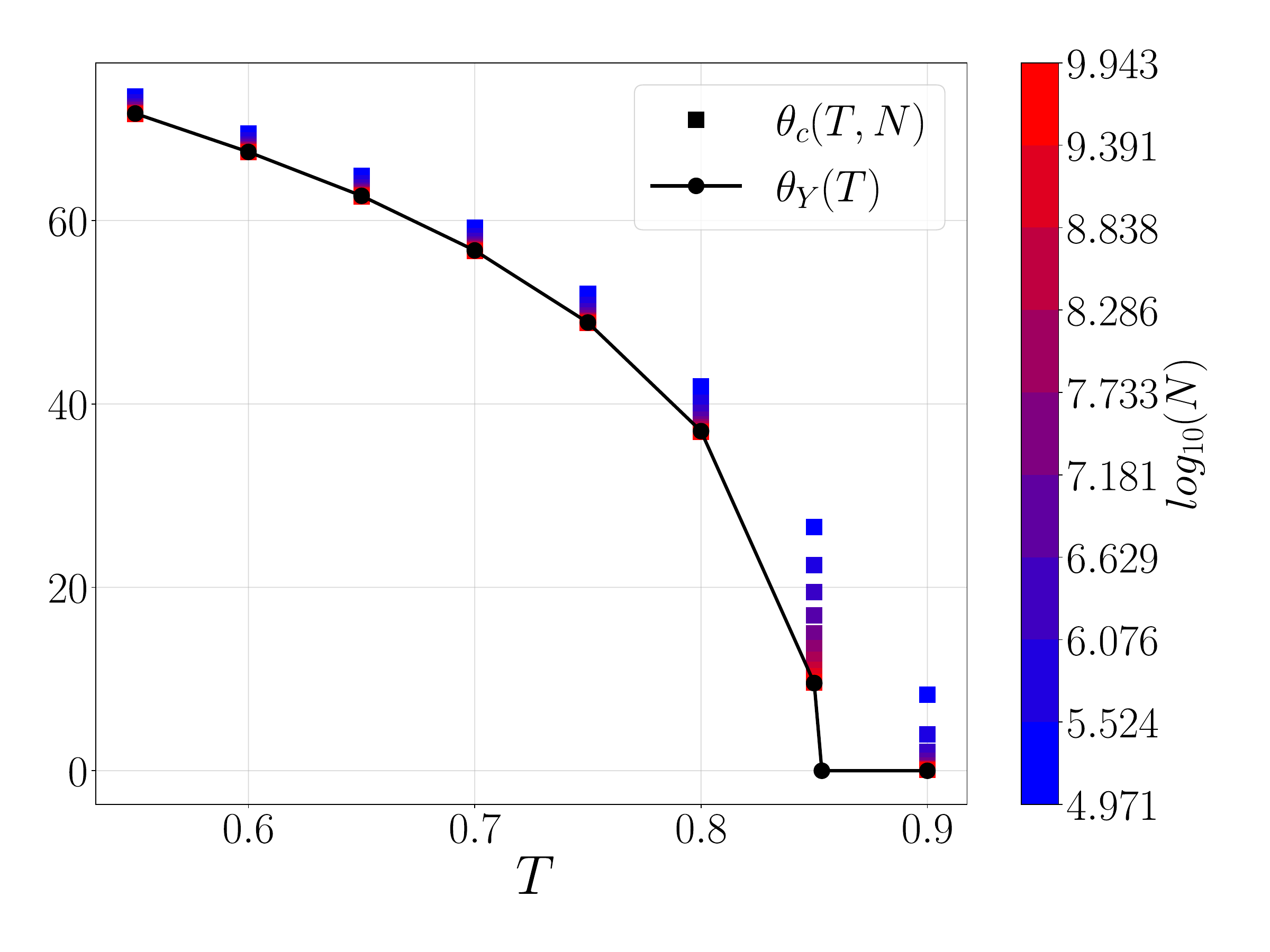}
    \caption{{Temperature and size dependence of  apparent contact angle $\theta_c(N,T)$. As $N$ is increased, $\theta_c$ tends to $\theta_Y$ at the same $T$. The dispersion of $\theta_c$ with $N$ increases with $T$, signifying the important role of binding potential in small drops. The Young angle $\theta_Y$ shows universal square-root scaling near $T_w$.}}
    \label{fig:theta_map}
\end{figure}
Present methodology offers principled means to obtain  apparent contact angles of adsorbed drops, across spatial scales. In Figure~\ref{fig:theta_map} we represent the $\theta_c (N, T)$-surface. It clearly shows that apparent contact angles tend to their respective Young values as drop sizes increase. 
At fixed temperature, $\theta_c$ decreases monotonically with $N$, converging smoothly to the macroscopic limit where $\theta_c$ approaches Young's contact angle $\theta_Y$. In contrast, at fixed droplet size, increasing $T$ reduces $\theta_c$ until  wetting takes place.
The figure follows the apparent contact angle universal square-root scaling law predicted by wetting theory \cite{Dietrich, peter_iop} near the wetting temperature $\theta_c \propto (T_w-T)^{1/2}$, provided the droplet is sufficiently large. Observe the increase of the contact angle dispersion with the drop size as temperature is increased. Smaller drops have larger apparent contact angle than bigger drops at the same temperature, because of the higher role played by the interface binding potential.

\section{Discussion}
\label{sec: discussion}
\subsection{Physics of the Learned Correction Terms}

\begin{figure}[h!]
    \centering
    \subfloat[]{\includegraphics[width=0.45\textwidth]{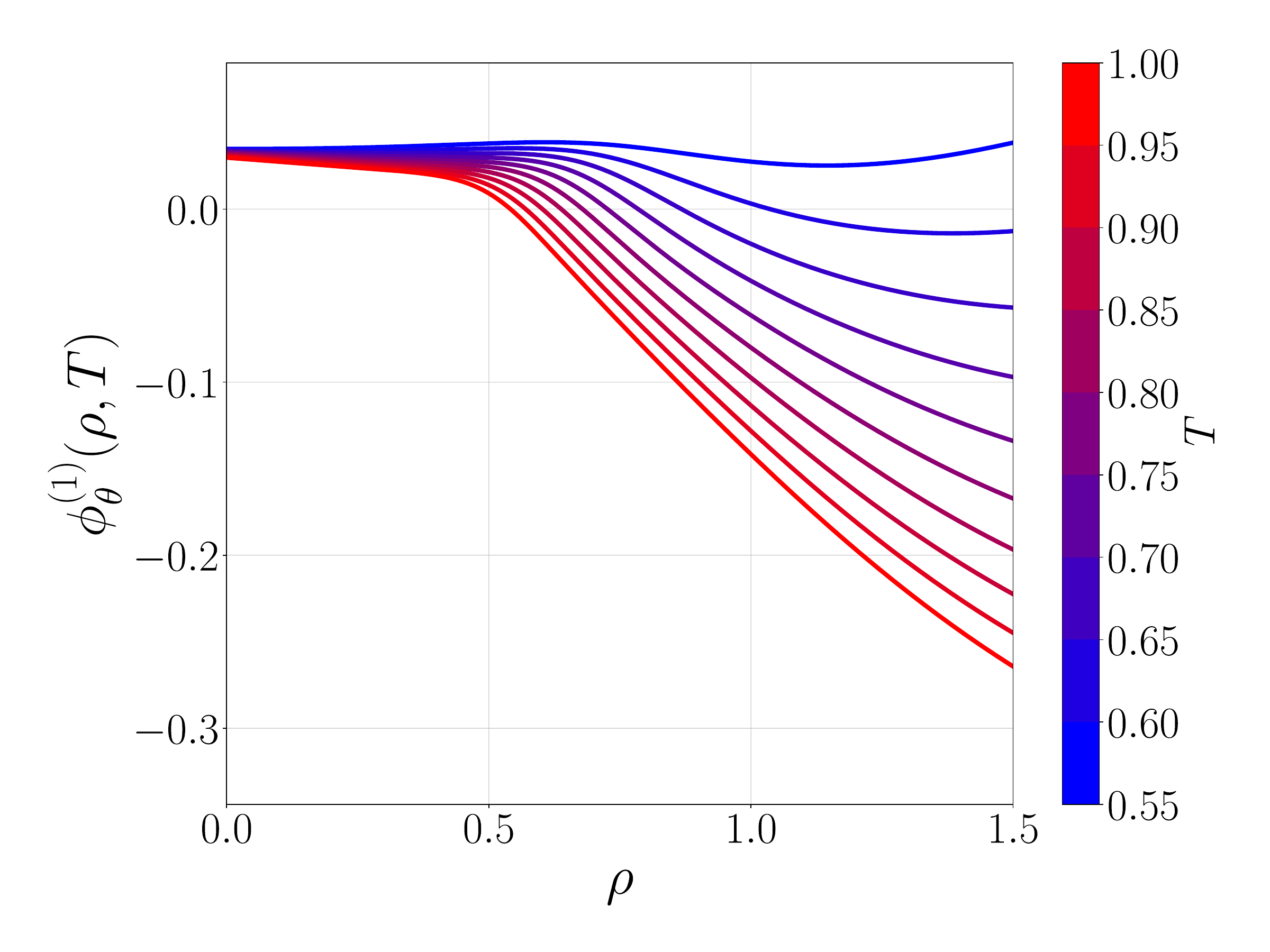}\label{fig:phi_1}}
    \hspace{0.05\textwidth}
    \subfloat[] {\includegraphics[width=0.45\textwidth]{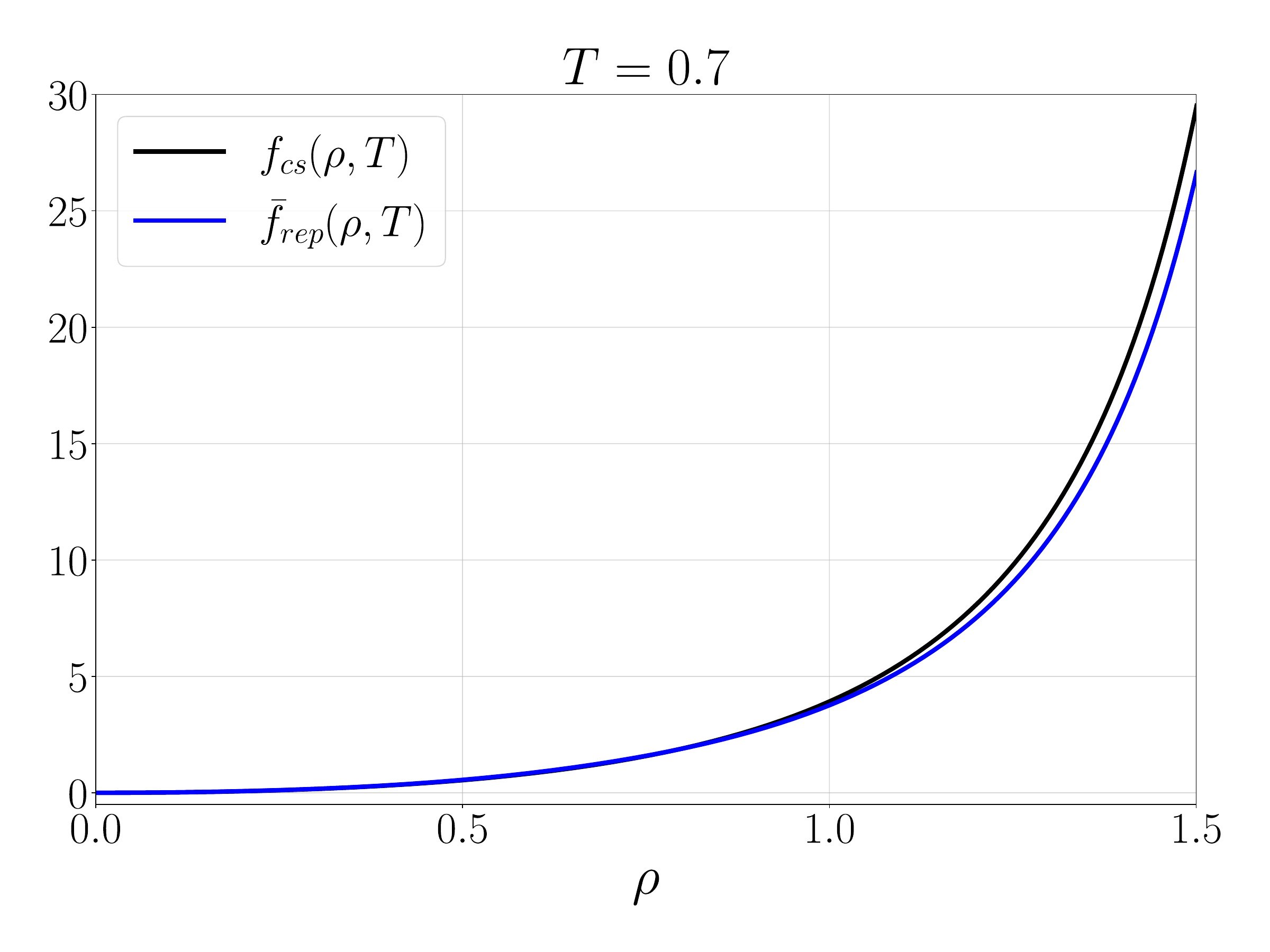}\label{fig:f_rep}}
    \caption{The physics captured by the EOS correction term $\phi^{(1)}_\theta(\rho,T)$. 
    (a) Isolines of  $\phi^{(1)}_\theta(\rho,T)$ at constant temperature.
    (b) Corrected EOS term $\bar f_{rep}(\rho,T)=(1+\phi^{(1)}_\theta(\rho,T))f_{cs}(\rho,T)$, superimposed against the CS baseline $f_{cs}(\rho, T)$ for $T=0.7$. }
\end{figure}

We begin by discussing each correction ${\phi_\theta}^{(k)}(\rho,T)$ in turn, revealing the relevant physics captured by our model. Figure~\ref{fig:phi_1} shows contours of the trained ${\phi_\theta}^{(1)}(\rho,T)$ at different $T$. Each contour is smooth. At a given $\rho$, ${\phi_\theta}^{(1)}(\rho,T)$ decreases with temperature. 
This is understandable, because we expect a reduction in the number of effective particle collisions as $T$ is lowered. 
Notice that, at sufficiently high density and temperature, the deviations from the CS equation of state become quasi linear functions of density.
The augmented excess Helmholtz free-energy density ${\bar f_{rep}}(\rho,T)$, shown in Figure~\ref{fig:f_rep}, deviates from the CS prediction at high densities, while converging to $f_{cs}$ at low densities, where it rapidly decays to zero. This indicates that the model correctly recovers the expected ideal-gas thermodynamic limit.

\begin{figure}[H]
    \centering
    \includegraphics[width=0.5\textwidth]{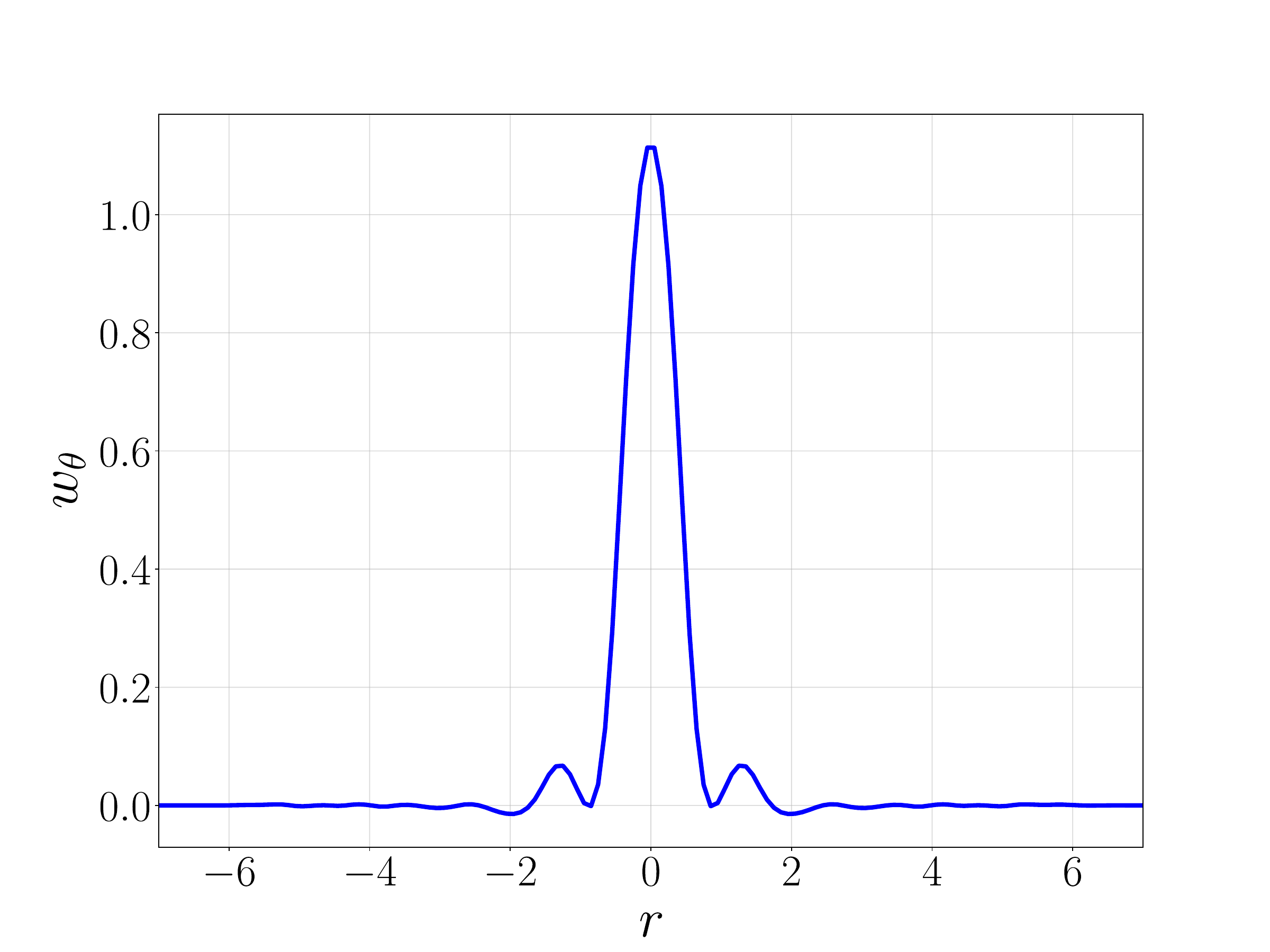}
    \caption{The learned WDA kernel $\phi^{(2)}_\theta(x)$, recovering the delta-peak at origin and oscillations at multiples of hard-sphere diameter.}
    \label{fig:phi2}
\end{figure}

The learned WDA kernel correction $\phi_{\theta}^{(2)}$, shown in Figure~\ref{fig:phi2}, exhibits the expected delta-like peak around $r=0$. Crucially, the model also captures rapidly decaying oscillations at approximately $r>1.5$. These account for the longer-range correlations caused by the particle packing effects. This behavior of our ML model is fully consistent with the sophisticated nonlocal theories of hard-sphere fluids, such as WDA and FMT~\cite{wda, tarazona2008_fmt_review}. 

\begin{figure}[h!]
    \centering
    \includegraphics[width=0.5\textwidth]{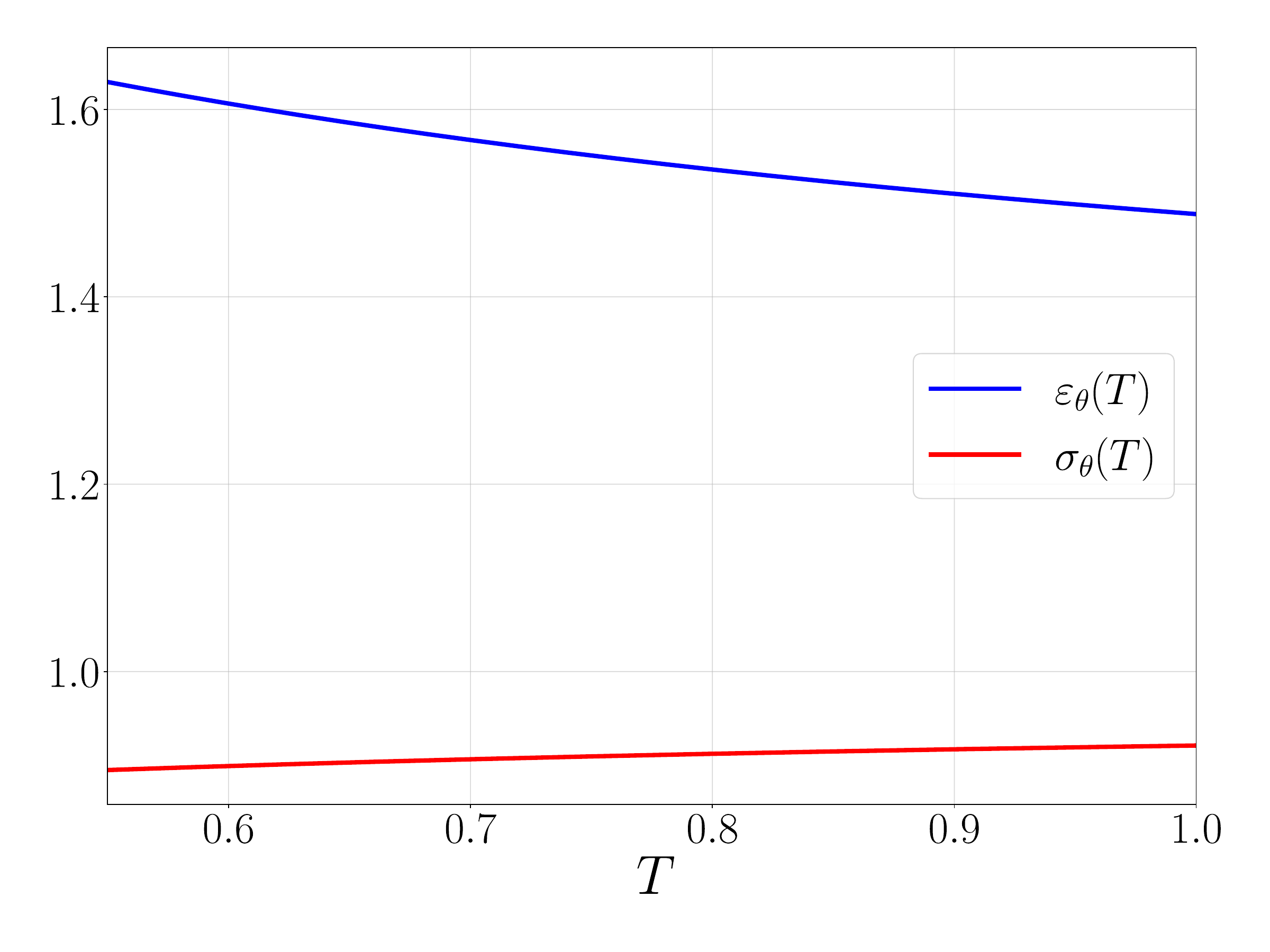}
    \caption{
    Temperature dependence of the calibrated effective Lennard–Jones parameters [$\bm \phi^{(3)}_\theta (T)$ in the main text]. The fitted attraction strength $\varepsilon_\theta(T)$ decreases with $T$, while the effective length scale $\sigma_\theta(T)$ - increases. The latter trend should be interpreted as a state-dependent effective parameter, rather than a physical hard-core diameter.
    }
    \label{fig:phi3}
\end{figure}

Figure~\ref{fig:phi3} shows the corrections $\bm \phi^{(3)}_\theta$~\eqref{eq:phi3} of the LJ parameters. The inferred temperature dependence should be interpreted with care. 
In particular, the monotonic decrease of the fitted attraction scale $\varepsilon_\theta(T)$ with increasing $T$ is consistent with a progressive weakening of effective cohesive interactions when unresolved degrees of freedom and many-body correlations are integrated out. 
By contrast, the concomitant increase of the fitted length scale $\sigma_\theta(T)$ does not admit a literal interpretation as a microscopic hard-core or excluded-volume diameter. Within the BH theory, the effective hard-sphere diameter decreases with temperature due to thermal softening of the repulsion. The opposite trend observed here indicates that $\sigma_\theta(T)$ should be regarded as an effective, state-dependent parameter rather than a physical particle size.

From this perspective, the joint evolution of $\varepsilon_\theta(T)$ and $\sigma_\theta(T)$ reflects a compensatory mechanism within the calibrated model, whereby systematic approximations in the underlying theory or closure are absorbed into temperature-dependent parameters. In particular, an increase in $\sigma_\theta$ acts to enhance excluded-volume effects at higher temperatures, offsetting deficiencies in the representation of short-range correlations, while a reduced $\varepsilon_\theta$ prevents overbinding. Such behavior is consistent with the interpretation of the fitted interaction as a coarse-grained, state-dependent effective potential rather than a fixed pairwise interaction.

\begin{figure}[h!]
    \centering
    \includegraphics[width=0.5\textwidth]{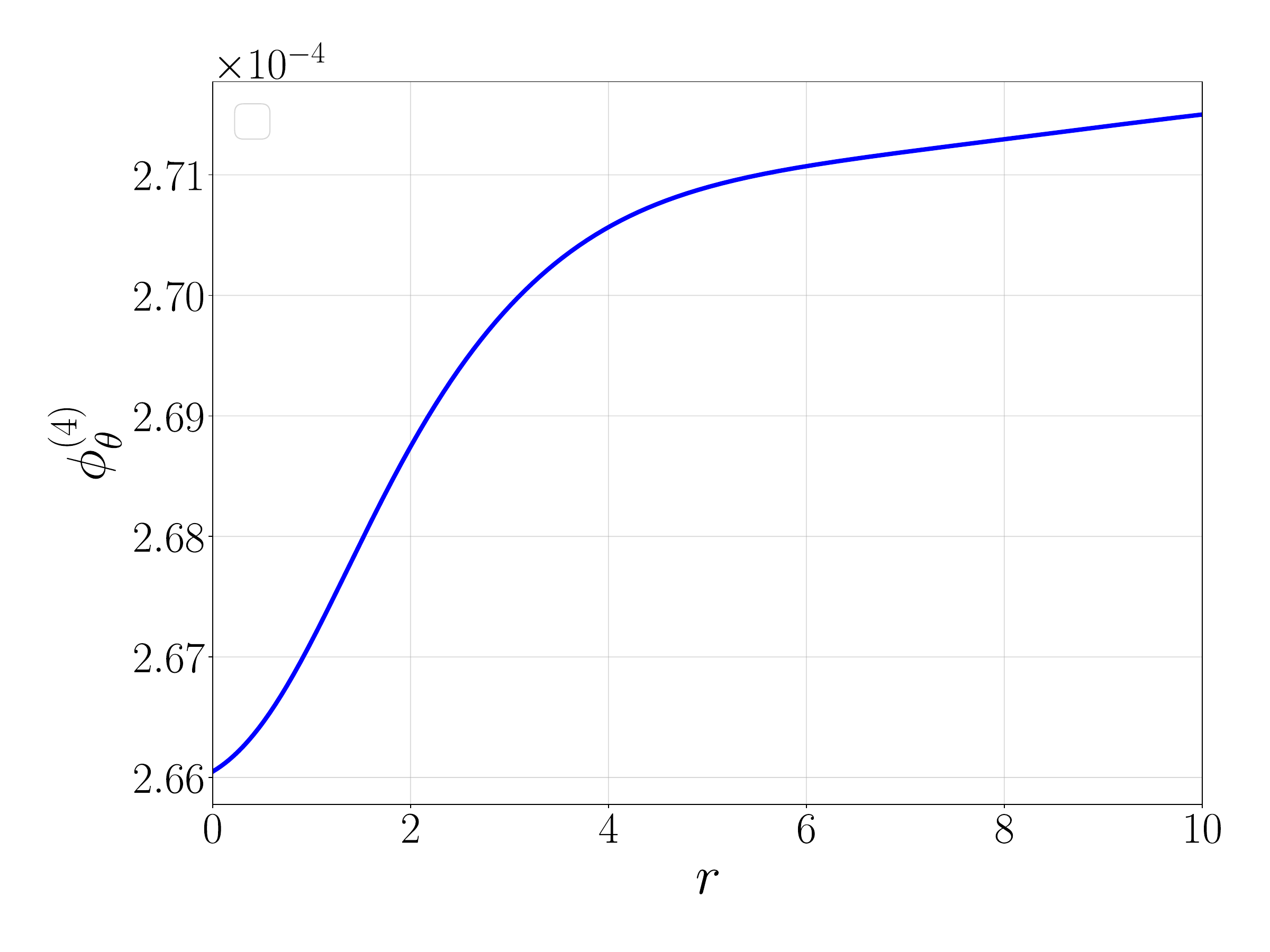}
    \caption{The learned correction to the attractive free energy, $g_{hs}(r)=1 + \phi^{(4)}_\theta (r)$ remains negligible, confirming the use of mean-field approximation.}
    \label{fig:phi4}
\end{figure}
Finally, we find that the model does not take advantage of the kernel in the attractive part, $g_{hs}(r)=1 + \phi^{(4)}_\theta (r)$. This is illustrated in Figure~\ref{fig:phi4}, where the learned correction has negligible scale. This indicates that the mean-field approximation of attractions is in practice appropriate, given that the reference hard-sphere fluid is adequately captured.

\subsection{Model Generalization Explained}

\begin{figure}[tb] 
\centering
\begin{minipage}{.7\textwidth}
\centering

\begin{tikzpicture}
\node[inner sep=0] (img) {\includegraphics[width=.9\textwidth]{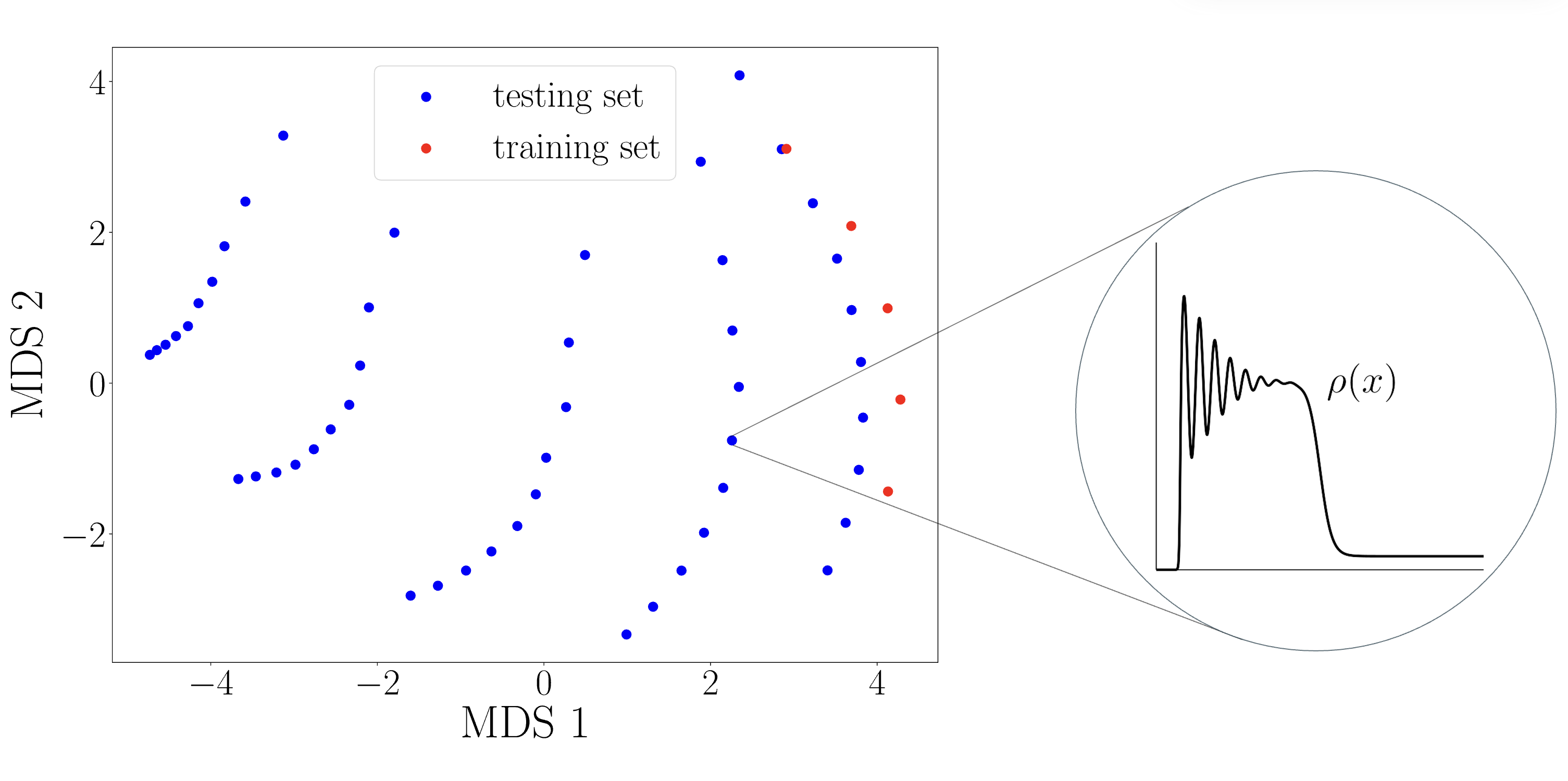}};
\node[anchor=east] at (img.west) {(a)  $\quad\mathcal{X}$};
\end{tikzpicture}

\vspace{1em}

\begin{tikzpicture}
\node[inner sep=0] (img) {\includegraphics[width=.9\textwidth]{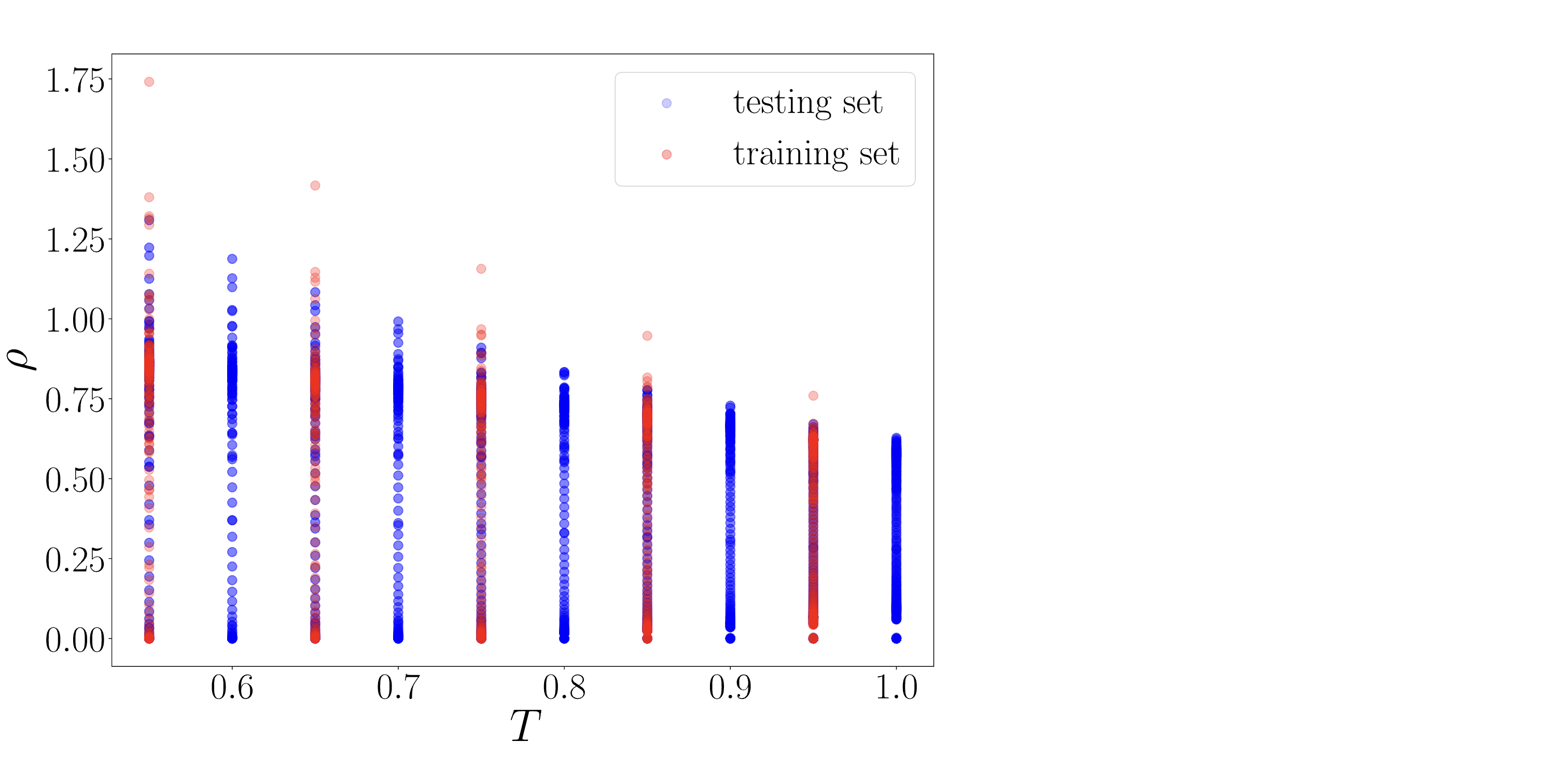}};
\node[anchor=east] at (img.west) {(b)  $\quad\mathcal{A}$};
\end{tikzpicture}

\end{minipage}

\caption{
    Illustration of the generalization capabilities of our ML model. 
    (a) shows the MDS representation of the five density profiles in the training set (red) and fifty profiles in the test set. Each point represents one profile $\rho^{(k)}(\,\cdot\,;\; T) \in \mathcal X$. The curve-like clusters correspond to the same $N$. Within each curve-like cluster, the points differ in $T$.
    (b) shows the same density profiles, discretized and flattened, plotted against their respective $T$, as points belonging to the low-dimensional $\mathcal{A}$. Each data point $\rho^{(k)}(\,\cdot\,;\; T) \in \mathcal X$ in (a), once projected to the points $P^{(k)}=\{(\rho^{(k)}(x_i ;\; T_j),T_j) \quad\forall x_i\in  D\}\in\mathcal A$ in (b), spans a continuous range of density values, which densely populate $\mathcal A$.
    }
\label{fig:low_dim_projection}
\end{figure}

The augmented models capture the temperature and density dependence of the corrections in a manner that is interpretable and coherent with molecular physics. 
The remarkable generalization we demonstrated in Sections~\ref{subsec:calibration_results} and~\ref{subsec:droplets} was achieved by the model, which was trained only on five adsorption density profiles, across different $T$. 
We can readily understand why by considering the effective dimensionality of our learning task. Although the training set contains only five density profiles, each individual profile $\rho^{(k)}(x;T) \in \mathcal{X}$ is a function which spans a continuous range of local density values in its spatial domain. 
By construction, these local densities constitute the actual input to our corrective neural models.
By mapping each spatial profile $\rho(\,\cdot\,;\;T)$ to the points $x \mapsto \bigl(\rho(x;T), T\bigr) \in \mathcal A \subset \mathbb{R}^2$, the data are embedded in the low-dimensional manifold $\mathcal{A}$. In a discretized representation, the points $\bigl(\rho^{(k)}(x_i;T_j),\,T_j\bigr)$ belonging to the function $\rho^{(k)}(\,\cdot\,;T_j)$ densely populate the region of $\mathcal{A}$ that spans from $\max[\rho^{(k)}]$ to $\min[\rho^{(k)}]$, as shown in Figure~\ref{fig:low_dim_projection}. 
All blue points belong to the testing set and correspond to the same wall potential. In contrast, the red points represent the five density profiles used for training, which were generated using a different wall potential.
Each of the curve-like clusters in Figure~\ref{fig:low_dim_projection}a corresponds to the same $N$. Within the same curve-like cluster, the points differ in $T$. 

Figure~\ref{fig:low_dim_projection} clearly shows that our NN is not attempting to extrapolate to a part of $\mathcal{X}$-space not covered by the training data. Instead, it interpolates inside a low-dimensional manifold, which is well covered by the training data. Although panel (a) suggests that the training set barely covers the functional space $\mathcal{X}$, panel (b) confirms that in the space that matters - $\mathcal A$ - our ML model is in fact interpolating. In this space, our limited training data are sufficient to capture the necessary physics for learning the cDFT appropriately. 
This explains the remarkable generalization achieved by our model when applying it to unseen $V_{ext}(x)$, $N$, $T$, etc, since the model interpolates within $\mathcal{A}$ even when it extrapolates in geometry and external potentials.

\section{Conclusion}\label{sec:4_discussion}

We presented a methodology to design and train hybrid physics-ML models for the excess Helmholtz free energy, introducing physically-motivated, but data-driven corrections to well-established theoretical models. We validated our principled approach on the benchmark single-component Lennard–Jones fluid. This is a well-understood system, which therefore provides a robust basis for assessing the performance of our methodology.
Our work demonstrates how machine-learned terms can be incorporated into the existing semi-empirical free-energy functionals, accurately predicting bulk and interfacial phenomena, including phase-coexistence curves, critical phenomena, adsorption profiles, surface tensions, and even shapes of nano-drops.
The HFE is independent of confining geometries and external potentials. Therefore we were able to learn correction to $F_\text{ex}$ using only small amounts of MD simulation data. This is because we carefully introduced the correct inductive bias, grounded in rigorous theoretical prior work.
Notably, these results were obtained despite training on a deliberately small dataset consisting of a limited number of one-dimensional planar adsorption profiles, in order to mimic realistic conditions of data scarcity. This efficiency arises from the strong inductive bias imposed by the underlying statistical-mechanical structure, which reduces the effective learning problem to a low-dimensional manifold of local thermodynamic states.
This demonstrates that our inductive bias substantially reduces the need for costly particle simulation data. In fact, once trained, our model fully captures the fluid physics, across a wide range of temperatures. The model generalizes to unseen geometries, external potentials and other thermodynamic conditions, without the need for additional training. This makes the corrections easily transferable. 
Another advantage of our hybrid physics-ML philosophy is its intrinsic interpretability. Avoiding black box models allows us to gain  understanding of the underlying phenomenology, helping rationalizing the downstream effects of the data-driven surrogates and grasping the physics that motivates them.
However, while the presented augmented model performs robustly across the examined temperature and size ranges, the corrective terms introduced here still represent a simplification with respect to the physics-prescribed contributions to $F_\text{ex}$. A more general approach could allow for the density-dependence of the WDA kernel and more complex baseline models. 
\\\\
Future work may extend the proposed framework along several complementary directions.
First, generalization to more complex fluids and to chemically or geometrically heterogeneous substrates would enable the investigation of more realistic wetting scenarios, including surface patterning, compositional effects, and multicomponent adsorption.
Second, the present formulation does not include uncertainty quantification~\cite{4_uq_active_learning}, which constitutes an important next step. Bayesian, physics-constrained inference frameworks~\cite{1_uq_baysesian_peter_andrew,10_uq_bayesian_Vext} provide a natural route to quantify uncertainty in learned corrections and in derived thermodynamic observables. This aspect is particularly critical for complex fluids, where experimental data are often scarce and high-fidelity molecular simulations are computationally prohibitive, making reliability assessment essential for data-driven models.
Third, coupling the augmented density functional with continuum-scale CFD solvers represents a promising avenue for multiscale modeling. Such integration would enable predictive simulations of macroscopic transport and wetting phenomena while consistently incorporating molecular-scale thermodynamic information.
\\\\
Overall, this work demonstrates that physics-informed learning, when carefully integrated into established statistical-mechanical frameworks, can substantially enhance the quantitative accuracy and applicability of classical density functional theory without sacrificing interpretability or physical consistency. Rather than replacing theory with black-box surrogates, the approach developed here illustrates how molecular simulation data can be systematically assimilated into continuum thermodynamic models, providing a scalable bridge between particle-level descriptions and macroscopic interfacial phenomena.

\section*{Acknowledgments}\label{sec:4_acknowledgments}
This research project has received funding from the European Union’s Horizon Europe research and innovation programme under the Marie Sklodowska-Curie Doctoral Network grant agreement no 101072578—Bridging Models at Different Scales To Design New Generation Fuel Cells for Electrified Mobility.

\appendix
\section*{Appendix}
\section{Molecular Dynamics Dataset for Planar Adsorption}
\label{appendix:adsorption_dataset}

The MD simulations were performed in the canonical $NVT$ ensemble. All quantities are reported in LJ reduced units. 
The system consists of an LJ fluid confined by a planar wall, perpendicular to the $x$-direction. 
Two datasets are constructed to train and assess the model performance across thermodynamic conditions and wall potentials. We consider the following sets of particle numbers and temperatures:
\begin{itemize}
    \item $\mathcal{N}=\{4000,\, 5000,\, 6000,\, 7000,\, 8000\}$,
    \item $\mathcal{T}=\{0.55,\, 0.6,\, 0.65,\, 0.7,\, 0.75,\, 0.8,\, 0.85,\, 0.9,\, 0.95,\, 1.0\}$.
\end{itemize}
Additionally, we consider two different wall potentials as in Eq.~\eqref{eq:lj93}, described by the parameters $\bm p_w$: 
\begin{itemize}
    \item $\mathcal{P}=\left\{\bm{p}^{(1)}_w=[1.2,\,1.2],\;\bm{p}^{(2)}_w=[1,\,2]\right\}$.
\end{itemize}

\begin{figure}[hbt!]
    \centering
    \includegraphics[width=0.75\textwidth]{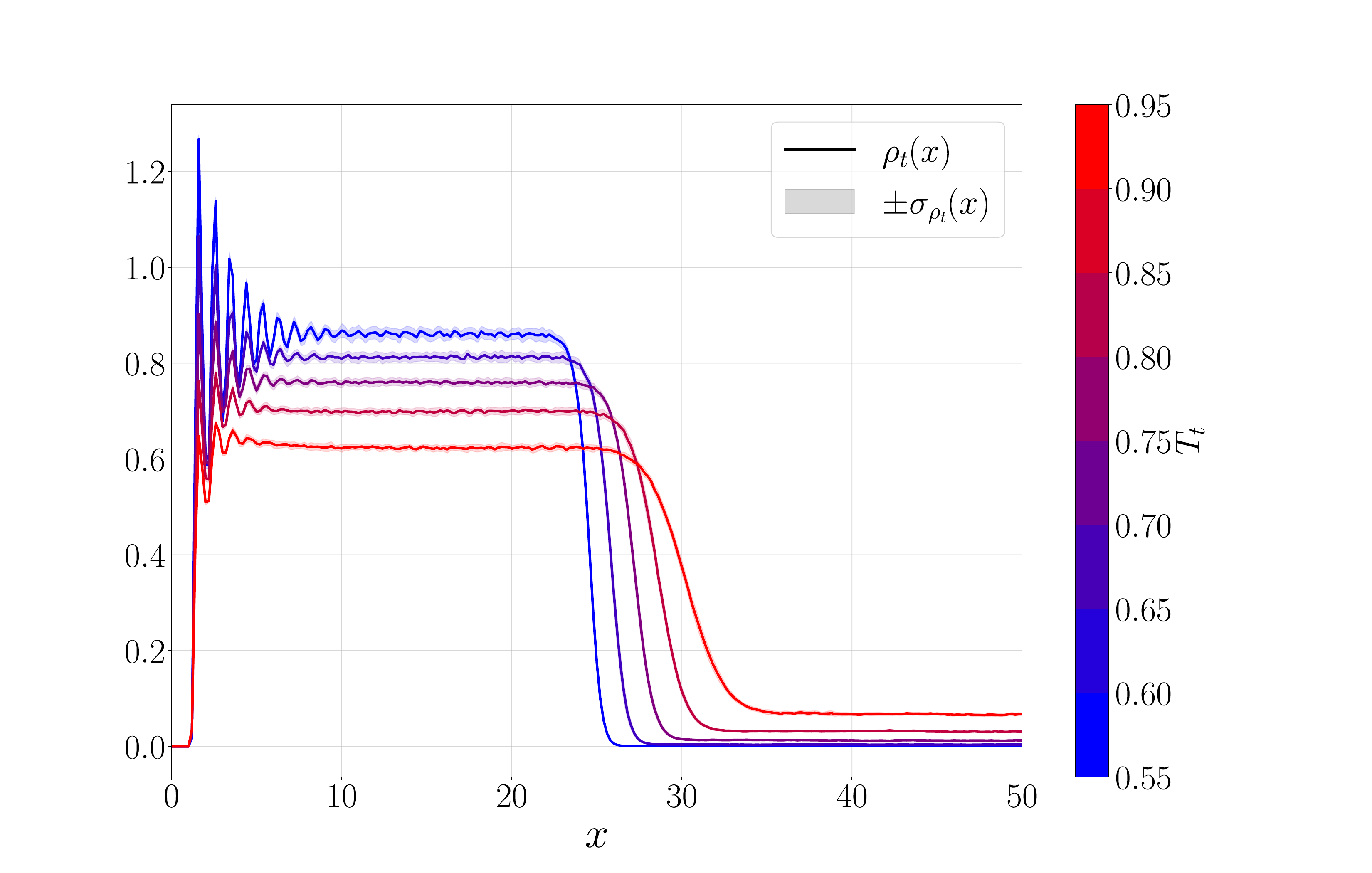}
    \caption{
        \textcolor{black}{
    Equilibrium density profiles corresponding to the five thermodynamic states in the training set $\mathcal D_t$. Simulations are performed with wall-potential parameters $\bm{p}^{(1)}_w=[1.2,\,1.2]$ and total number of particles $N=8000$. The density profiles, corresponding to $T=0.55,\, 0.65,\, 0.75,\, 0.85,\, 0.95$, are shown as solid curves $\rho_t(x)$ with shaded confidence bands $\rho_t(x)\pm  \sigma_{\rho_t}(x)$, where $\sigma_{\rho_t}(x)$is the standard deviation of the sampled density at position $x$.}}
    \label{fig:training_set}
\end{figure}

\paragraph{Training Dataset.}
The training dataset samples a deliberately restricted thermodynamic subspace in order to emulate realistic data-scarcity conditions. 
It is generated using the wall potential $\mathbf{p}_w = \mathbf{p}^{(1)}_w$ over the following thermodynamic states:
\begin{itemize}
    \item $\mathcal{N}_t = \{8000\}$,
    \item $\mathcal{T}_t = \{0.55,\, 0.65,\, 0.75,\, 0.85,\, 0.95\}$,
\end{itemize}
Each entry contains the thermodynamic state, wall parameters, and the equilibrium density profile obtained by binning the corresponding MD-simulated trajectories:
\[
\mathcal{D}_t = \{\mathcal{D}^{(j)}_t\}_{j=1}^{M_t},
\qquad
\mathcal{D}^{(j)}_t =
\left\{
T^{(j)}_t,\,
N^{(j)}_t,\,
\bm{p}^{(1)}_w,\,
\rho^{(j)}_t(x)
\right\},
\]
with a total of $M_t = 5$ data points. The density profiles in the training set are shown in Figure~\ref{fig:training_set}, along with their corresponding confidence bands.

\paragraph{Reference Dataset for Validation.}
The reference dataset $\mathcal{D}_\text{ref}$ is used to evaluate model performance. 
It comprises all combinations $(N^{(j)},T^{(j)})\in(\mathcal{N}\times\mathcal{T})$ and is generated using the wall potential $\mathbf{p}_w = \mathbf{p}^{(2)}_w$.
Analogously, each dataset entry contains the thermodynamic state, wall parameters, and the equilibrium density profile obtained by binning the corresponding MD-simulated trajectories:
\[
\mathcal{D}_\text{ref} = \{\mathcal{D}^{(j)}_\text{ref}\}_{j=1}^{M_\text{ref}},
\qquad
\mathcal{D}^{(j)}_\text{ref} =
\left\{
T^{(j)}_\text{ref},\,
N^{(j)}_\text{ref},\,
\bm{p}^{(2)}_w,\,
\rho^{(j)}_\text{ref}(x)
\right\},
\]
with a total of $M_\text{ref}=50$ simulations.

Together, these datasets enable a systematic assessment of model robustness and generalization capabilities across thermodynamic conditions, and transferability with respect to wall--fluid external potentials.

\section{Adjoint Method for Operator Optimization}
\label{appendix:adjoint_training}

The adjoint method provides an efficient way to compute the gradient of a general objective ${J}$ with respect to a set of trainable parameters ${\bm \theta}$ when the state function $\rho(x)$, with $x\in D$, satisfies the residual operator ${R}([\rho],\,\cdot\, ;\;{\bm \theta}) = 0$.
This is achieved by introducing an auxiliary variable, the adjoint field, $\lambda$, which enforces the residual constraint.

The Lagrangian is defined as
\begin{equation}
    {L}([\rho({\bm \theta})],x)
    = {J}([\rho],x)
    + \int_D \lambda(x,y)\, {R}([\rho],y;{\bm \theta}) \, dy,
\end{equation}
where ${J}([\rho],\,\cdot\,)$ is an arbitrary objective operator, and ${R}([\rho],\,\cdot\,;\;{\bm \theta})$ enforces the governing equation.
Differentiating the Lagrangian with respect to the parameters ${\bm \theta}$ gives:
\begin{equation}
\label{eq:BalanceEquations}
\begin{aligned}
    \partial_{\bm \theta} {L}
    &=
    \int_D \frac{\delta {J}(x)}{\delta \rho(z)}\, \partial_{\bm \theta} \rho(z) \, dz
    + \iint_{D\times D} \lambda(x,y)
        \frac{\delta {R}([\rho], y; {\bm \theta})}{\delta \rho(z)}
        \, \partial_{\bm \theta} \rho(z)\, dz\, dy
    + \int_D \lambda(x,y)
     \partial_{{\bm \theta}} {R}([\rho], y; {\bm \theta})\, dy.
\end{aligned}
\end{equation}

Grouping the terms that multiply $\partial_{\bm \theta}\rho$ gives
\begin{equation}
\label{eq:GroupedVariation}
    \partial_{\bm \theta} {L}
    =
    \int_D \left[
        \frac{\delta {J}(x)}{\delta \rho(z)}
        + \int_D \lambda(x,y)\,
        \frac{\delta {R}([\rho], y;{\bm \theta})}{\delta \rho(z)}\, dy
    \right]
    \partial_{\bm \theta} \rho(z)\, dz
    + \int_D \lambda(x,y)
        \partial_{{\bm \theta}} {R}([\rho], y; {\bm \theta})\, dy.
\end{equation}

The first integral in Eq.~\eqref{eq:GroupedVariation} can be eliminated by requiring its integrand to vanish for any admissible variation $\partial_{\bm \theta}\rho$, which leads to the {adjoint equation}:
\begin{equation}
\label{eq:AdjointEquation}
    \frac{\delta {J}([\rho],x)}{\delta \rho(z)}
    + \int_D \lambda(x,y)\,
      \frac{\delta {R}([\rho], y;{\bm \theta})}{\delta \rho(z)}\, dy
    = 0.
\end{equation}
Solving Eq.~\eqref{eq:AdjointEquation} provides the adjoint field $\lambda(x,y)$. 
The total sensitivity of the objective with respect to the parameters follows directly from the remaining term in Eq.~\eqref{eq:GroupedVariation}:
\begin{equation}
\label{eq:Sensitivity}
    \partial_{\bm \theta} {J}
    =
    \int_D \lambda(x,y)\,
    \partial_{{\bm \theta}} {R}([\rho], y; {\bm \theta}) \, dy.
\end{equation}
This compact expression provides the exact gradient of the objective $J$ given the residual constraint.
In the specific case of the grand-canonical Euler--Lagrange Eq.~\eqref{eq:EL-muVT}, $R=h_{gc}$. The functional derivative of Eq.~\eqref{eq:EL-muVT} gives:
\begin{equation}
\begin{aligned}
\label{eq:Rgc_def}
    \frac{\delta h_{\mathrm{gc}}([\rho], y; \bm \theta )}{\delta \rho (z)}
    &=
    \delta(y-z)
    - \beta \big( h_{\mathrm{gc}}([\rho], y; \bm \theta ) - \rho(y) \big)
      \frac{\delta^2 F_{\mathrm{ex}}([\rho];\bm \theta )}{\delta \rho(y)\delta \rho(z)}.
\end{aligned}
\end{equation}
For the canonical formulation, the normalization constraint introduces an additional correction term that ensures conservation of the total number of particles. The resulting functional derivatives is
\begin{equation}
\label{eq:Rc_def_canonical}
\begin{aligned}
    \frac{\delta h_{\mathrm{c}}([\rho], y; \bm \theta )}{\delta \rho (z)}
    &=
    \delta(y-z)
    + \beta\, h_{0,c}([\rho],y; \bm\theta)
    \int_{D}
    \left[
        \delta(x-y)
        - \frac{h_{0,c}([\rho],x; \bm\theta)}{N}
    \right]
    \frac{\delta^{2}F_{\mathrm{ex}}([\rho];\bm \theta )}{\delta\rho(z)\delta\rho(x)}\,dx,
\end{aligned}
\end{equation}
where \(h_{0,c} = \rho - h_{\mathrm{c}}\).  
Solving the adjoint equation~\eqref{eq:AdjointEquation} for $\lambda$ 
and substituting it into Eq.~\eqref{eq:Sensitivity} 
returns the gradient of the $J$ with respect to the parameters $\bm \theta$. 
In practice, the functional derivatives ${\delta F_{\mathrm{ex}}}/{\delta\rho}$ and ${\delta^{2} F_{\mathrm{ex}}}/{\delta\rho^{2}}$ can be evaluated automatically within the computational graph, 
through nested automatic differentiation of the eHFE model, while the residual derivatives $\partial R/\partial\bm{\theta}$ are computed directly by backpropagation. 
This procedure enables efficient and fully differentiable training of the physics-constrained augmented functional model.

\printbibliography
\end{document}